\PassOptionsToPackage{ colorlinks=true, citecolor=darkred, urlcolor=darkred, pdfborder={0 0 0} }{hyperref}

\documentclass[aps,prd,notitlepage,showpacs,nofootinbib,superscriptaddress,floatfix]{revtex4-2}
\pdfoutput=1

\usepackage[usenames,dvipsnames]{color}
\usepackage{graphicx}
\usepackage{subcaption}
\usepackage{bm}
\usepackage{bbold}
\usepackage{latexsym}
\usepackage{amsthm}
\usepackage{mathtools}
\usepackage{amsmath}
\usepackage{xfrac}
\usepackage{wrapfig}
\usepackage{amssymb}
\usepackage{cancel}
\usepackage{multirow}
\usepackage{rotating}
\usepackage{hhline}
\usepackage{float}
\usepackage{fancybox}
\usepackage[utf8]{inputenc}
\usepackage{booktabs}
\usepackage{array}
\usepackage{pdfcomment}
\usepackage[tooltip]{acro}
\usepackage{xspace}
\usepackage{soul}
\usepackage{framed}
\newcolumntype{C}[1]{>{\centering\let\newline\\\arraybackslash\hspace{0pt}}m{#1}}

\usepackage[normalem]{ulem}
\usepackage{cleveref}
\usepackage{afterpage}

\setul{0.5ex}{0.3ex}
\setulcolor{blue}

\usepackage{braket}
\definecolor{linkcolor}{rgb}{0,0,0.5}
\allowdisplaybreaks[2]

\definecolor{greenLinks}{rgb}{0, 0.6, 0}
\definecolor{blueLinks}{rgb}{0, 0, 0.6}
\definecolor{redLinks}{rgb}{0.6, 0, 0}
\definecolor{tempText}{rgb}{0.55, 0.10,0.67}
\definecolor{eprintLinks}{rgb}{0.4, 0.4, 0.4}
\definecolor{journalLinks}{rgb}{0.6, 0, 0}

\definecolor{mightnightblue}{RGB}{25,25,112}

\definecolor{brown}{rgb}{0.59, 0.29, 0.0}

\definecolor{darkred}{rgb}{0.6,0,0}

\def\lsim{\mathrel{\rlap{\lower4pt\hbox{\hskip1pt$\sim$}} \raise1pt\hbox{$<$}}}
\def\gsim{\mathrel{\rlap{\lower4pt\hbox{\hskip1pt$\sim$}} \raise1pt\hbox{$>$}}}

\def\U1s{$\mathrm{U_{1}^{(a)}\otimes U_{1}^{(b)}\otimes U_{1}^{(c)}\otimes U_{1}^{(d)}\otimes U_{1}^{(e)}}$ }
\def\3211{$\mathrm{SU(3) \times SU(2)_L \times U(1)_R \times U(1)_{B-L}}$ }
\def\321{$\mathrm{SU(3) \times SU(2) \times U(1)}$ }
\def\422{$\mathrm{SU(4) \times SU(2) \times SU(2)_R}$ }

\DeclareMathOperator{\diag}{diag}
\DeclareMathOperator{\tr}{Tr}
\DeclareMathOperator{\im}{Im}
\DeclareMathOperator{\re}{Re}

\acsetup{pdfcomments/use=true}
\DeclareAcronym{BSM}{ short = BSM , long = beyond the standard model, pdfcomment = beyond the standard model }
\DeclareAcronym{BU}{ short = BU, long = bottom-up, pdfcomment = bottom-up }
\DeclareAcronym{CG}{ short = CG, long = Clebsch--Gordan, pdfcomment = Clebsch--Gordan }
\DeclareAcronym{EFT}{ short = EFT, long = effective field theory, pdfcomment = effective field theory }
\DeclareAcronym{GUT}{ short = GUT, long = grand unified theory, pdfcomment = grand unified theory }
\DeclareAcronym{IR}{ short = IR, long = infrared, pdfcomment = infrared }
\DeclareAcronym{irrep}{ short = irrep, long = irreducible representation, pdfcomment = irreducible representation }
\DeclareAcronym{MSSM}{ short = MSSM, long = minimal supersymmetric standard model, pdfcomment = minimal supersymmetric standard model }
\DeclareAcronym{OPM}{ short = OPM, long = one-parameter modular model, pdfcomment = one-parameter modular model }
\DeclareAcronym{QFT}{ short = QFT, long = quantum field theory, pdfcomment = quantum field theory }
\DeclareAcronym{SM}{ short = SM, long = standard model of particle physics, pdfcomment = standard model of particle physics }
\DeclareAcronym{RG}{ short = RG, long = renormalization group, pdfcomment = renormalization group }
\DeclareAcronym{RGE}{ short = RGE, long = renormalization group equation, pdfcomment = renormalization group equation }
\DeclareAcronym{SUSY}{ short = SUSY, long = supersymmetry, pdfcomment = supersymmetry }
\DeclareAcronym{TD}{ short = TD, long = top-down, pdfcomment = top-down }
\DeclareAcronym{UV}{ short = UV, long = ultraviolet, pdfcomment = ultraviolet }
\DeclareAcronym{VEV}{ short = VEV, long = vacuum expectation value, pdfcomment = vacuum expectation value }
\DeclareAcronym{VVMF}{ short = VVMF, long = vector-valued modular form, pdfcomment = vector-valued modular form }

\newcommand{\AddrAHEP}{  Instituto de Física Corpuscular (IFIC), Universidad de Valencia-CSIC,\\ Paterna (Valencia) E-46980, Spain}

\definecolor{darkgreen}{rgb}{0,0.6,0}

\definecolor{darkpurple}{rgb}{0.45,0.0,0.65}

\begin{document}

\title{\boldmath Fermion mass relations in one-parameter modular models}

\author{Salvador Centelles Chuli\'{a}}
\email{salcen@ific.uv.es}
\affiliation{\AddrAHEP}
\author{Xueqi Li}
\email{xueqi.li@uci.edu}
\affiliation{Department of Physics and Astronomy, University of California, Irvine, CA 92697-4575 USA}
\author{Xiang-Gan Liu}
\email{xianggal@uci.edu}
\affiliation{Department of Physics and Astronomy, University of California, Irvine, CA 92697-4575 USA}
\affiliation{Instituto de F\'isica, Universidad Nacional Aut\'onoma de M\'exico, Cd.\ de M\'exico C.P.\ 04510, M\'exico}
\author{Omar Medina}
\email{omar.medina@ific.uv.es}
\affiliation{\AddrAHEP}
\author{J. T. Penedo}
\email{jpenedo@roma3.infn.it}
\affiliation{INFN Sezione di Roma Tre, Via della Vasca Navale 84, 00146, Roma, Italy}
%%%%%%%%%%%%%%%%%%%%%%%%%%%%%
\begin{abstract}
\vskip 0mm
Modular flavour symmetries provide a possible organizing principle for the Standard Model Yukawa sector, by replacing generic couplings with a potentially small number of modular forms controlled by a single complex modulus. We study the extreme limit of this idea: \acp{OPM}, in which each charged-fermion mass matrix is fixed by a single modular invariant contraction. We develop a systematic method to construct such models, showing that the \ac{OPM} requirement is already highly constraining at the level of possible fermion hierarchies. In a concrete realization, the charged-lepton and down-quark sectors are controlled by the common modulus, leading to exact mass relations at the flavour scale,
\vskip -3mm
\[
m_s^5 = 2\sqrt{2}\,m_d^3m_b^2,
\qquad
m_\mu^3 = \sqrt{2}\,m_e m_\tau^2,
\qquad
m_s^2m_\tau = \sqrt{2}\,m_e m_b^2.
\]
We show that, once renormalization-group evolution and selective supersymmetric threshold effects are included, these high-scale relations can be made compatible with low-energy charged-fermion data. Our results provide a working proof of principle for \acp{OPM} and point towards a possible route to the flavour puzzle through highly constrained constructions.
\end{abstract}
%%%%%%%%%%%%%%%%%%%%%%%%%%%%%

\maketitle

%%%%%%%%%%%%%%%%%%%%%%%%%%%%%
\section{Introduction}
%%%%%%%%%%%%%%%%%%%%%%%%%%%%%

The Standard Model (SM) allows for an extraordinarily economical description of fundamental interactions, based on the gauge symmetry principle. However, this economy is lost in its Yukawa sector, which encodes fermion masses, mixing and CP violation (CPV), and where most of the theory's free parameters reside. Explaining the peculiarities of the observed flavour structures, namely the marked mass hierarchies and contrasting mixing patterns (i.e. large for leptons and small for quarks), constitutes the flavour puzzle and remains an open challenge in theoretical particle physics.

An organizing principle in the flavour sector is desired: it would work orthogonally to the gauge principle, unifying different particle generations into multiplets. Non-Abelian flavour (or horizontal) symmetries provide such a framework, postulating the invariance of the action under the transformations of a symmetry group $G_f$, suitably broken by the \acp{VEV} of scalar fields, known as flavons (see e.g.~\cite{Altarelli:2010gt,Ishimori:2010au,King:2014nza,Petcov:2017ggy,Feruglio:2019ybq,Ding:2024ozt} for reviews). The modular invariance generalization of the traditional flavour approach~\cite{Feruglio:2017spp} (see~\cite{Kobayashi:2023zzc,Ding:2023htn} for reviews) focuses instead on a minimal symmetry-breaking sector, described by \emph{a single complex field}: the modulus $\tau$, with $\im\tau>0$. Within this string-inspired, supersymmetric framework, modular forms $Y(\tau)$ play the role of Yukawa couplings which are significantly constrained. In particular, the holomorphicity of the superpotential restricts the space of available forms, making this setup potentially very predictive.\footnote{\label{foot:SUSYbreaking}We work with the minimal modular-invariant Kähler potential. As usual in modular flavour models, non-minimal Kähler corrections are allowed by the symmetry and may reduce predictivity by introducing additional parameters~\cite{Chen:2019ewa}. We regard the minimal choice as part of the definition of the framework.} Moreover, model-independent mass relations may emerge from the symmetry structure of the flavour group and of the modular forms in modular-invariant models containing a small number of parameters, see e.g.~\cite{Chen:2023mwt}. Taken to the extreme, one could in principle build a modular model where each Yukawa matrix results from \emph{a single invariant contraction}. Within such a one-parameter modular model, each fermion mass matrix is thus fully determined by $\tau$ and an overall coefficient.

In this work, we investigate the conditions under which \acp{OPM} may be viable. A first step towards building an \ac{OPM} involves accounting for the hierarchical fermion spectrum, which already places significant constraints on the possible values of the modulus \ac{VEV}. Note that a generic \ac{VEV} of $\tau$ fully breaks the non-linearly realized modular invariance. Instead, at the special values $\tau_\text{sym} = i,\, \omega= e^{2\pi i/3},\,i\infty$ some residual $\mathbb{Z}_n$ ($n\geq 2$) symmetry is preserved and can be exploited. Indeed, in the vicinity of these fixed points, the residual symmetry is slightly broken (but linearly realized) and Yukawa couplings can be expanded as power series in $\epsilon \sim |\tau - \tau_\text{sym}|$, providing a natural origin of the fermion mass hierarchies~\cite{Feruglio:2021dte,Novichkov:2021evw}. This mechanism, which has recently~\cite{Carducci:2026shs} been dubbed \emph{modular proximity-induced hierarchies}, has been used to derive ``golden''-type mass relations in models relying on a limited number of parameters in Ref.~\cite{Chen:2023mwt}. Finally, we stress that the number of possible \acp{OPM} one can construct may be limited, due to the structure of the modular group $\mathrm{SL}(2,\mathbb{Z})$. As a result, obtaining the observed charged-fermion mass hierarchies from an \ac{OPM} is highly non-trivial.

\vskip 2mm
This paper is organized as follows. In~\Cref{sec:oneParaIntro} we introduce the modular-invariant framework and the conditions under which a single invariant contraction gives rise to an \ac{OPM} Yukawa sector with suitable mass hierarchies. In~\Cref{sec:models} we use these criteria to search for possible \acp{OPM}. This search first singles out $\Delta(96)$ and the related group $\Delta(384)$, where four inequivalent hierarchy patterns H$_2$\,--\,H$_5$ are obtained. Two of these hierarchies, H$_2$ and H$_5$, are naturally close to the observed charged-lepton and down-quark spectra. Assigning them to the corresponding sectors in an explicit double \ac{OPM} then fixes the high-scale mass relations studied in the rest of the paper. In~\Cref{sec:phenomenology} we confront these relations with low-energy data, through renormalization-group running and finite threshold corrections. In~\Cref{sec:quark-flavour-hints} we discuss a possible one-parameter up-quark hierarchy. We conclude in~\Cref{sec:conclusions}. An explicit up-quark completion and further results are given in the appendices.

%%%%%%%%%%%%%%%%%%%%%%%%%%%%%
\section{Framework}
\label{sec:oneParaIntro}
%%%%%%%%%%%%%%%%%%%%%%%%%%%%%

We consider a simple setup in flavour modular symmetry. We work in global $\mathcal{N}=1$ supersymmetry, with matter chiral superfields $\psi^i$ acquiring their masses from the Yukawa interactions in the superpotential, characterized by cubic terms of the type
\begin{equation} \label{eq:W}
    \mathcal W \,\supset\, Y_{ijk}(\tau)\,\psi_i\psi_j\psi_k\,,
\end{equation}
where the $Y_{ijk}$ are the components of a \ac{VVMF} (or modular form multiplet). These are holomorphic functions that transform under the modular group $\mathrm{SL}(2,\mathbb Z)$ as
\begin{equation} \label{eq:VVMFTransform}
    Y^{(k_Y)}(\tau) \,\,\xmapsto{\gamma}\,\, Y^{(k_Y)}(\gamma \tau) \,=\, (c\,\tau + d)^{k_Y} \rho(\gamma)\, Y^{(k_Y)}(\tau)\,, \qquad \text{with }
    \gamma =
    \begin{pmatrix}
        a & b \\
        c & d
    \end{pmatrix}
    \in \mathrm{SL}(2,\mathbb Z)\,,
\end{equation}
where $\rho$ is a representation of $\mathrm{SL}(2,\mathbb Z)$, $k_Y$ is known as the weight of the modular form, and $\tau$ is the modulus, which transforms as
\begin{equation}
    \tau \,\,\xmapsto{\gamma}\,\,\gamma \tau \,=\, \frac{a \, \tau + b}{c \, \tau + d}\,.
\end{equation}
Recall that $\mathrm{SL}(2,\mathbb{Z})$ is generated by the matrices $S = \left(\begin{smallmatrix} 0 & 1 \\-1 & 0 \end{smallmatrix}\right)$ and $T = \left(\begin{smallmatrix} 1 & 1 \\ 0 & 1 \end{smallmatrix}\right)$. We focus on the case where the representation $\rho$ has finite image; in such a case, the \acp{VVMF} are those of the finite modular group $\mathrm{SL}(2,\mathbb Z)/\ker \rho$. A matter field transforms under the modular group as
\begin{equation}
    \psi\,\,\xmapsto{\gamma}\,\, (c\,\tau + d)^{-k_\psi} \rho_\psi(\gamma)\, \psi\,,
\end{equation}
where $k_\psi$ is the modular weight of the matter field $\psi$ and $\rho_\psi$ is a (unitary) representation of the finite modular group. In global supersymmetry, the superpotential is invariant under modular transformations. Then, together with~\cref{eq:VVMFTransform}, one finds that, in each superpotential term, the sum of the matter field modular weights must equal the weight of the modular form, $k_Y = k_{\psi_i} + k_{\psi_j} + k_{\psi_k}$. For relatively low weights, there are only a few, or even a single, modular form(s) available, strongly constraining flavour structures and observables.

In building modular-invariant flavour models, we consider a minimal form for the Kähler potential,
\begin{equation}
    \mathcal K =  - h\, \Lambda_\tau^2 \log (-i \tau + i \bar\tau) + \sum_i  (-i \tau + i \bar\tau)^{-k_{\psi_i}} |\psi_i|^2\,,
\end{equation}
where $h>0$ and $\Lambda_\tau$ has a mass dimension of 1. In a given charged fermion sector, assuming a single Higgs field $H$ per sector ($H_u$ or $H_d$), the modular-invariant Yukawa term can be expanded as
\begin{equation} \label{eq:Wsum}
    \mathcal W \supset \sum_i \alpha_i \left(Y_{\mathbf{r}_i}^{(k_i)}(\tau)\,\psi^c \psi \right)_\mathbf{1} H \;,
\end{equation}
where the $\alpha_i$ are constant parameters and $(\cdots)_\mathbf{1}$ denotes the contraction of flavour indices into a trivial singlet. Each Higgs doublet is taken to transform as a trivial singlet in flavour space with modular weight $k_H= 0$, without loss of generality.\footnote{In particular, non-trivial transformation properties of a Higgs superfield can be absorbed by the weights and \acp{irrep} of matter fields within the bilinear.} The modular weight $k_i$ is then equal to the sum of the weights of the parts of $\psi^c$ and $\psi$ participating in the contraction. The sum in~\cref{eq:Wsum} runs over all possible modular form representations and weights whose contraction yields a trivial singlet.

We are interested in the extreme one-parameter case, in which there is only one $\alpha_i$ for each sector, denoted $\alpha_f$ ($f =u,d,e$).\footnote{While similar considerations may be applied to the neutrino sector, the corresponding analysis differs since light neutrinos are not necessarily hierarchical and the lightest state may be massless at leading order.} In such a scenario, there is exactly one possible contraction for a given $k = k_\psi + k_{\psi^c}$, i.e.~the above sum must reduce to a single term. To satisfy this requirement, two conditions must be met:
\begin{enumerate}
\item among the representations $\mathbf{r}_i$ under which the \acp{VVMF} $Y_{\mathbf{r}_i}^{(k)}$ transform, only one representation, $\mathbf{r}$, can contract with $\psi^c$ and $\psi$ to produce a trivial (non-vanishing) singlet;\label{item:1r} \item  there is only one modular form of weight $k$ furnishing the representation $\mathbf{r}$;
\label{item:1form}
\end{enumerate}
If either of these conditions is not met, more than one contraction will be present.

For a model with three generations of massive fermions, all three generations must participate in the contraction so that each generation acquires a mass. Together with the condition in~\cref{item:1r}, this implies that the fermions must transform as a triplet of the finite modular group. Otherwise, either some generations remain massless because they do not participate in the contraction, or there is more than one contraction, since each singlet or doublet must be contracted separately. We therefore focus only on the case in which $\psi$ and $\psi^c$ transform as triplets, since only in this case can an \ac{OPM} be obtained.

To satisfy the condition in~\cref{item:1form}, it is useful to consider only modular forms of the lowest weight. At low modular weight, there is typically only one modular form available. In this work, we restrict our search to weights 1 and 2, which are the lowest weights for modular forms in most finite modular groups.

\begin{table}[t!]
\centering
\renewcommand{\arraystretch}{1.2}
\begin{tabular}{cccc} \toprule
 \quad{ }Type\quad{ }  & Max rank at $\tau_\text{sym}$
 & \quad{ }Leading spectrum\quad{ } & Possible asymptotic regions \\ \midrule
 I & $1$ & $c_1\,\epsilon^p:c_2\,\epsilon^q:1 $ & $\tau\simeq i\infty$ \\
 II & $2$ & $c_1\,\epsilon^p:c_2:1 $ & $\tau\simeq i\infty,~ i$ \\
 III & $3$ & $c_1:c_2:1$ &  $\tau\simeq i\infty, ~i ,~\omega$ \\ \bottomrule
\end{tabular}
\caption{Classification of mass spectra in \acp{OPM}, where $p,q\neq 0$ and the constants $c_i$ follow from the expansion coefficients of modular forms and \ac{CG} coefficients of the finite modular group.}
\label{tab:threeTypeMirical}
\end{table}

\vskip 2mm
As discussed in the introduction, we further narrow our focus by considering only models predicting (non-zero) hierarchical masses. Indeed, one can naturally obtain hierarchical fermion spectra by expanding around zeros in modular space~\cite{Feruglio:2021dte,Novichkov:2021evw}. More precisely, the determinant of the mass matrix, which is a one-dimensional modular form (see e.g.~Appendix A of~\cite{Penedo:2024gtb}), vanishes at these values of $\tau$. The transformation properties and the zeroes of modular determinants have been discussed in detail in~\cite{Chen:2025tby}. One can show that, within the fundamental domain of $\mathrm{SL}(2,\mathbb Z)$, the determinant modular forms of weight below 12 can vanish only at three points: $\tau = i\infty$, $i$, and $\omega = e^{{2\pi i}/{3}}$, known as the critical or fixed points. By expanding mass matrix elements around these points in terms of a small deviation parameter $\epsilon \sim |\tau - \tau_\text{sym}|$, one can categorize the resulting mass hierarchies into three types, as presented in \Cref{tab:threeTypeMirical}.\footnote{Notably, in modular flavour models, a single mass matrix can predict multiple distinct mass (or mixing angle) patterns depending on the location of the modulus \ac{VEV} in different asymptotic regions of moduli space. This is itself a rather interesting feature compared to traditional flavour symmetry models.}

Note that some spectra are not available in a given asymptotic region. For instance, the spectrum $c_1\,\epsilon^2 : c_2\,\epsilon: 1$ is not attainable if $\tau \simeq \omega$ in this context. This follows from the stringent requirement that $\psi$ and $\psi^c$ are irreducible triplets. Indeed, as noted in~\cite{Feruglio:2023mii}, in this irreducible case one has
\begin{itemize}
\item $\psi\,\,\xmapsto{ST}\,\, \diag(1,\omega,\omega^2)\, \psi$ in an appropriate $ST$-diagonal basis; \item $\psi\,\,\xmapsto{S}\,\, i^{k_S} \diag(1,-1,-1) \psi$ in an appropriate $S$-diagonal basis, where $k_S$ is an integer in the cases of interest.
\end{itemize}
This implies that, at the symmetric point $\tau = \omega$ one expects a spectrum of the type $c_1:c_2:1$, while for $\tau = i$ one expects either $c_1:c_2:1$ or $0:c_2:1$ (or a fully massless spectrum). At nearby values of $\tau$, massless fermions are generically lifted by an appropriate power of $\epsilon$.

For spectra of type I, the hierarchy is naturally given by powers of the small parameter $\epsilon$. For type III, the hierarchy would instead arise purely from the coefficients of the modular forms and group tensor products, an intriguing possibility~\cite{Chen:2025tby}. Type II lies between these two cases. In this work, we focus on spectra of type I, leading us to consider large $\im \tau$. As such, the appropriate expansion parameter $\epsilon$ is an appropriate power of $|q|$, with $q\equiv e^{2\pi i\tau}$ and $q\to 0$ as $\tau \to i \infty$. These criteria define the search performed in the next section.

%%%%%%%%%%%%%%%%%%%%%%
\section{One-parameter modular models}
\label{sec:models}
%%%%%%%%%%%%%%%%%%%%%%

An \ac{OPM} is defined, in part, by the finite modular group $\mathrm{SL}(2,\mathbb Z)/\ker \rho$, which must admit at least one triplet \ac{irrep}. A list of the non-Abelian finite modular groups with order $<78$ has been given in Ref.~\cite{Liu:2021gwa}. Up to (and including) order 100, one finds the following finite modular groups admitting triplet \acp{irrep}:
\begin{equation} \label{eq:groups}
\begin{array}{llll}
 [12, 3] \simeq A_4                    \,\, (1)\,,
&[24, 3] \simeq T'                     \,\, (1)\,,
&[24,12] \simeq S_4                    \,\, (2)\,,
&[24,13] \simeq A_4 \times \mathbb{Z}_2\,\, (2)\,,\\[2mm]
 [48,28] \simeq 2O                     \,\, (2)\,,
&[48,29] \simeq GL(2,3)                \,\, (2)\,,\quad{}
&[48,30] \simeq S_4'                   \,\, (4)\,,
&[48,31] \simeq A_4 \times \mathbb{Z}_4\,\, (4)\,,\\[2mm]
 [48,32] \simeq T'  \times \mathbb{Z}_2\,\, (2)\,,
&[48,33] \simeq T'  \circ  \mathbb{Z}_4\,\, (2)\,,
&\boldsymbol{[60, 5] \simeq A_5 \,\, (2)}\,,
&[72,42] \simeq S_4 \times \mathbb{Z}_3\,\, (6)\,,\\[2mm]
 [72,44] \simeq A_4 \times S_3         \,\, (2)\,,\quad{}
&\boldsymbol{[96,64] \simeq \Delta(96) \,\, (6)}\,,
&[96,66] \simeq T'  \rtimes\mathbb{Z}_4\,\, (4)\,,\quad{}
&[96,69] \simeq T'  \times \mathbb{Z}_4\,\, (4)\,,\\[2mm]
 \multicolumn{2}{l}{[96,72] \simeq ((\mathbb{Z}_4 \times \mathbb{Z}_4) \rtimes \mathbb{Z}_2) \rtimes \mathbb{Z}_3
                                       \,\, (2)\,,}
\end{array}
\end{equation}
where each group is first identified by its \textsc{GAP} Id~\cite{GAP4} in brackets. The number in parentheses counts the number of distinct triplet \acp{irrep} available.

Focusing on \emph{hierarchical} spectra of type I, i.e. of the type~$c_1\,\epsilon^p:c_2\,\epsilon^q:1$ with $p\neq q$, we require that the mass matrix obtained from the contraction of the pair of triplets $\psi^c \otimes \psi$ has at most rank 1, in the symmetric limit $\tau \to i \infty$. Note that a residual $\mathbb{Z}_N^T$ symmetry is recovered in that limit, where $N$ is the order of $\rho(T)$. From the decomposition of the triplets $\psi$ and $\psi^c$ under $\mathbb{Z}_N^T$, one can directly infer the rank of the mass matrix and how its zeroes are lifted for $\epsilon \neq 0$ in a $T$-diagonal basis~\cite{Novichkov:2021evw}. This selects only two groups from~\cref{eq:groups}, highlighted in bold, as well as a small number of pairs of triplets as promising \ac{OPM} ingredients.

An $A_5$-based \ac{OPM} would represent an attractive possibility. Unfortunately, as already noted in section 3.3.2 of Ref.~\cite{Novichkov:2021evw}, this potential \ac{OPM} predicts a massless fermion in the SUSY limit, since the determinant of the mass matrix vanishes identically for any value of $\tau$.\footnote{We are interested in realizing fermion mass hierarchies in a unified way, requiring a non-vanishing determinant when $\epsilon \neq 0$. See however Ref.~\cite{Feruglio:2021dte} for a discussion on lifting the massless fermion via SUSY-breaking effects or dimension-six operators.} This can be understood from the fact that such a weight-6 determinant, with a zero at $\tau = i \infty$, transforms as a trivial singlet of $\mathrm{SL}(2,\mathbb{Z})$. Following~\cite{Chen:2025tby}, one can see that these properties force it to be constantly zero, even if $\epsilon \neq 0$.

\vskip 2mm

Motivated by the above discussion, one may consider the only promising group at order $\leq 100$, namely $\Delta(96) \simeq [96,64]$, which is a member of the $\Delta(6n^2)$ family of groups, with $n=4$, and is a subgroup of the finite modular group $\Gamma_8$~\cite{deAdelhartToorop:2011re}. We find that it is possible to build an \ac{OPM} based on $\Delta(96)$, albeit leading to a spectrum which may only realistically describe the charged-lepton sector (see below). Motivated by this partial success, in what follows we mostly concentrate on exploring a larger group, $\Delta(384) \simeq [384,568] \subset \Gamma_{16}$, which contains $\Delta(96)$ as a subgroup and is also a finite modular group. It is, in fact, the first element of the $\Delta(6n^2)$ family ($n=8$) that contains $\Delta(96)$ as a proper subgroup. While this larger group is chosen to illustrate a richer set of \ac{OPM} predictions, note that it is not the result of a systematic search, in contrast with the exhaustive scan of~\cref{eq:groups}. Such a search would need to consider 42 additional finite modular groups of order $<384$ that admit triplet \acp{irrep}. The presentations of the considered groups read:
\begin{equation}
\begin{aligned}
\Delta(96)&\,=\,\langle S,T ~\big|~ S^2=(ST)^3=(ST^{-1}ST)^3=T^8=1 \rangle\,,\\
\Delta(384)&\,=\,\langle S,T ~\big|~ S^2=(ST)^3=(ST^{-1}ST)^3=T^{16}=1 \rangle \,,
\end{aligned}
\end{equation}
so that $N = 8$ for $\Delta(96)$ and $N = 16$ for $\Delta(384)$. By explicit construction, we find that \acp{OPM} that can be obtained using $\Delta(96) \simeq \Delta(384)/(\mathbb Z_2 \times \mathbb Z_2)$ also arise in the search based on $\Delta(384)$.

\vskip 2mm

The group $\Delta(384)$ admits as \acp{irrep}: two singlets $\mathbf{1},\mathbf{1}'$, one doublet $\mathbf{2}$, fourteen triplets $\mathbf{3}_{0,1,\dots,13}$, and seven sextets $\mathbf{6}_{0,1,\dots,6}$ (all even representations, i.e.~$\rho(S)^2=\mathbb{1}$). To obtain \acp{OPM}, as illustrated in~\Cref{sec:oneParaIntro}, we focus on the lowest non-trivial weight, which is $k = 2$ for this group. There are only a few \acp{VVMF} at this weight, namely
\begin{equation} \label{eq:weight2_multiplets}
\begin{aligned}
k=2:\quad
&Y^{(2)}_{\mathbf{2}}(\tau),~ Y^{(2)}_{\mathbf{3}_0}(\tau),~ Y^{(2)}_{\mathbf{3}_3}(\tau),~ Y^{(2)}_{\mathbf{3}_5}(\tau),~
Y^{(2)}_{\mathbf{3}_7}(\tau),\\
&Y^{(2)}_{\mathbf{6}_0}(\tau),~ Y^{(2)}_{\mathbf{6}_2}(\tau),~ Y^{(2)}_{\mathbf{6}_3}(\tau),~
Y^{(2)}_{\mathbf{6}_4}(\tau),~ Y^{(2)}_{\mathbf{6}_5}(\tau)\,,
\end{aligned}
\end{equation}
in our convention. Using these forms and scanning over triplet tensor products $\mathbf{3}_i \otimes \mathbf{3}_j$, we are able to obtain several \acp{OPM} leading to spectra of the desired type for large $\im\tau$.

We summarize our results in~\Cref{tab:H12345}, which lists the corresponding hierarchical spectra, at leading order in an expansion in the small variable $\epsilon$, defined as~\cite{Novichkov:2021evw}
\begin{equation} \label{eq:epsilon}
\epsilon \equiv |q|^{1/N} = e^{-2\pi \im \tau / N}\ll 1\,.
\end{equation}
In particular, we have $\epsilon = |q|^{1/8}$ for $\Delta(96)$ and $\epsilon = |q|^{1/16}$ for $\Delta(384)$. We find a total of 10 triplet pairs that lead to \acp{OPM} based on the $\Delta(96)$ and $\Delta(384)$ finite modular groups. The complete list of triplet pairs and the corresponding hierarchy patterns is collected in~\Cref{app:A}. These are pairwise physically equivalent, corresponding to each of the 5 spectra in~\Cref{tab:H12345}. Moreover, one sees that the $\Delta(96)$-based H$_1$ \acp{OPM} are equivalent to the $\Delta(384)$-based H$_2$ ones, cf.~\cref{eq:epsilon}. In the case of H$_3$, the mass matrix vanishes in the symmetric limit and a global $\epsilon$ has been factored out when displaying the corresponding spectrum. Recall that the coefficients in the spectra are fixed by the structure of the modular forms and by the \ac{CG} coefficients of the finite modular group. In short, \emph{there are no additional parameters to adjust}, and ratios between fermion masses are fully determined in the limit of unbroken SUSY.

\begin{table}[t!]
\centering
\renewcommand{\arraystretch}{1.2}
\begin{tabular}{cccc} \toprule
 \quad{ } Hierarchy label\quad{ }  & \quad{ }Leading spectrum $m_1:m_2:m_3$\quad{ } & \quad{ }Finite modular group\quad{ } & \qquad{ }$p/q$\qquad{ }\\ \midrule
H$_1$ & $4\sqrt{2}\,\epsilon^{3} : 2\,\epsilon : 1$ & $\Delta(96)$ & $3$ \\
H$_2$ & $4\sqrt{2}\,\epsilon^{6} : 2\,\epsilon^{2} : 1$ & $\Delta(384)$ & $3$\\
H$_3$ & $2\sqrt{2}\,\epsilon^{4} : \sqrt{2}\,\epsilon : 1$ & $\Delta(384)$ & $4$\\
H$_4$ & $8\,\epsilon^{7} : \sqrt{2}\,\epsilon : 1$ & $\Delta(384)$ & $7$\\
H$_5$ & $4\,\epsilon^{5} : 2\sqrt{2}\,\epsilon^{3} : 1$ & $\Delta(384)$ & $5/3$\\ \bottomrule
\end{tabular}
\caption{Mass hierarchies obtained within \acp{OPM} and the associated flavour groups. Note that H$_1$ is physically equivalent to H$_2$.}
\label{tab:H12345}
\end{table}

Curiously, for all the found \acp{OPM} (cf.~\Cref{tab:H12345}), the coefficients $c_1$ and $c_2$ describing the leading-order spectrum, in the notation $c_1\,\epsilon^p:c_2\,\epsilon^q:1$, additionally obey $\sqrt{2}\,c_1 = c_2^{p/q}$, leading to the meta-relation
\begin{equation} \label{eq:meta}
    \left(\frac{m_2}{m_3}\right)^{\frac{p}{q}} \,=\, \sqrt{2} \, \frac{m_1}{m_3}\,,
\end{equation}
which is independent of $\epsilon$ and relates fermion masses within a given sector. It only depends on the ratio $p/q$, which characterizes each hierarchy and is shown explicitly in the last column of~\Cref{tab:H12345}.

\vskip 2mm

In what follows, we present in more detail two representative \acp{OPM}, corresponding to spectra of the type H$_2$ and H$_5$. As we will see, these may simultaneously describe, \emph{with a common value of $\epsilon$}, the charged-lepton and down-quark sectors, respectively. Consider assignments such that
\begin{equation} \label{eq:assign}
\begin{aligned}
 k_{e^c}+k_{L}=2             \,,\qquad{ }\,
(\rho_{e^c},\rho_{L}) &\sim (\mathbf{3}_1,\mathbf{3}_2)\,,
\\
 k_{d^c}+k_{Q}=2             \,,\qquad{ }
(\rho_{d^c},\rho_{Q}) &\sim (\mathbf{3}_7,\mathbf{3}_8)\,,
\end{aligned}
\end{equation}
where $Q$ and $L$ denote the quark and lepton doublet superfields, while $d^c$ and $e^c$ refer to the down-quark and charged-lepton singlet superfields. Here, the triplet \acp{irrep} can be uniquely identified by:
\begin{equation} \label{eq:3id}
\frac{1}{2\pi i}\log(\rho_\mathbf{r})(T)\,\,\,\,\text{mod}\,\mathbb{Z}\, =\,
\begin{cases}
\diag\left(\frac{1}{2},\,\frac{1}{4},\,\frac{3}{4}\right)\,,&\text{for }\mathbf{r} = \mathbf{3}_1\,,\\
\diag\left(\frac{3}{4},\,\frac{7}{8},\,\frac{3}{8}\right)\,,&\text{for }\mathbf{r} = \mathbf{3}_2\,,\\
\diag\left(\frac{3}{8},\,\frac{9}{16},\,\frac{1}{16}\right)\,,&\text{for }\mathbf{r} = \mathbf{3}_7\,,\\
\diag\left(\frac{1}{8},\,\frac{3}{4},\,\frac{5}{8}\right)\,,&\text{for }\mathbf{r} = \mathbf{3}_8\,.
\end{cases}
\end{equation}
To check that one obtains a potentially viable \ac{OPM}, one must i) verify that only one contraction is possible at the selected modular weight, and ii) confirm that the resulting Yukawa matrix has a determinant which is not constantly zero. This is the case for the two models presented here, for which the triplet contractions read:
\begin{equation}
\mathbf{3}_1\otimes\mathbf{3}_2
\,=\, \mathbf{3}_3 \oplus \mathbf{6}_0 = \overline{\mathbf{3}}_2 \oplus \overline{\mathbf{6}}_0 \,,
\qquad\quad
\mathbf{3}_7\otimes\mathbf{3}_8
\,=\, \mathbf{3}_5 \oplus \mathbf{6}_4 = \overline{\mathbf{3}}_4 \oplus \overline{\mathbf{6}}_4 \,,
\end{equation}
where a bar indicates the conjugate \ac{irrep}, $\mathbf{r} \otimes \overline{\mathbf{r}} \supset \mathbf{1}$, showing explicitly which representation is required to obtain an invariant. An inspection of~\cref{eq:weight2_multiplets} indicates that only one contraction (that with a sextet form) is possible for each product. The modular form multiplets of interest, $Y^{(2)}_{\mathbf{6}_0}$ and $Y^{(2)}_{\mathbf{6}_4}$, can be expanded as
\begin{align}
 Y^{(2)}_{\mathbf{6}_0}(\tau)=\begin{pmatrix}
  -8\, q^{3/8} (1 + 3\, q +\dots) \\
  2\, q^{1/8} (1 + 13\, q +\dots) \\
4\, q^{1/2}(-1 + 4 \, q +\dots  )  \\
1 - 8\, q^2 +\dots \\
-4\, q^{5/8} (3 + 7\, q +\dots)\\
   -16\, q^{7/8} (1 + 3\, q +\dots)
 \end{pmatrix}
 \,\,&\sim\,\,
 \begin{pmatrix}
-8\, \epsilon^6 \\
2\, \epsilon^2 \\
-4\, \epsilon^8\\
1 \\
-12\, \epsilon^{10}\\
-16\, \epsilon^{14}
\end{pmatrix}
\,+\, \mathcal{O}(\epsilon^{18})
 \,,
 \\[2mm]
 Y^{(2)}_{\mathbf{6}_4}(\tau)=\begin{pmatrix}
  2\sqrt{2}\, q^{11/16} (5 + 10\, q +\dots) \\
  2\sqrt{2}\, q^{3/16} (1 + 9\, q +\dots) \\
 -2\,q^{1/2}(1 - 6\, q +\dots  )  \\
 -1 + 6\, q +\dots \\
4\sqrt{2}\, q^{5/16} (1 + 4\, q +\dots)\\
-4\sqrt{2}\, q^{13/16} (3 + 7\, q +\dots)
 \end{pmatrix}
 \,\,&\sim\,\,
\begin{pmatrix}
10\sqrt{2}\, \epsilon^{11} \\
2\sqrt{2}\, \epsilon^3 \\
-2\, \epsilon^8\\
-1 \\
4\sqrt{2}\, \epsilon^5\\
-12\sqrt{2}\, \epsilon^{13}
\end{pmatrix}
\,+\, \mathcal{O}(\epsilon^{16})
 \,,
\end{align}
with $q = e^{2\pi i\tau}$ and $\epsilon = |q|^{1/16}$. Analytical expressions for these forms as well as more complete $q$-expansions are provided in~\Cref{app:A}. Already at this stage one sees that the approximate $\mathbb{Z}_N^T$ symmetry strongly suppresses corrections to leading order results, which arise at rather high orders in $\epsilon$.

The modular-invariant Yukawa terms are simply
\begin{equation}
\mathcal{W}_e \,=\, \alpha_e\left(Y^{(2)}_{\mathbf{6}_0}(\tau)\, e^c L\right)_{\mathbf{1}} H_d\,,\qquad
\mathcal{W}_d \,=\, \alpha_d\left(Y^{(2)}_{\mathbf{6}_4}(\tau)\, d^c Q\right)_{\mathbf{1}} H_d\,,
\end{equation}
where $\alpha_{d,e}$ are overall factors that, together with $v_d = \langle H_d^0\rangle$, will set the corresponding mass scales. Indeed, after electroweak symmetry breaking, the charged-lepton and down-quark mass matrices explicitly read
\begin{align}
M_e(\tau) &\,=\, \alpha_e\,v_d
\begin{pmatrix}
0 & \sqrt{2}\,Y^{(2)}_{\mathbf{6}_0,5} & -\sqrt{2}\,Y^{(2)}_{\mathbf{6}_0,2} \\
\sqrt{2}\,Y^{(2)}_{\mathbf{6}_0,4} & Y^{(2)}_{\mathbf{6}_0,6} & Y^{(2)}_{\mathbf{6}_0,1} \\
-\sqrt{2}\,Y^{(2)}_{\mathbf{6}_0,3} & -Y^{(2)}_{\mathbf{6}_0,1} & -Y^{(2)}_{\mathbf{6}_0,6}
\end{pmatrix}
\, \sim \,
\begin{pmatrix}
0 & -12\sqrt{2}\,\epsilon^{10} & -2\sqrt{2}\,\epsilon^2 \\
\sqrt{2} & -16\,\epsilon^{14} & -8\,\epsilon^6 \\
4\sqrt{2}\,\epsilon^8 & 8\,\epsilon^6 & 16\,\epsilon^{14}
\end{pmatrix}
\,+\, \mathcal{O}(\epsilon^{18})
\,,
\label{eq:Me}\\[2mm]
M_d(\tau) &\,=\, \alpha_d\,v_d
\begin{pmatrix}
-\sqrt{2}\,Y^{(2)}_{\mathbf{6}_4,3} & 0 & \sqrt{2}\,Y^{(2)}_{\mathbf{6}_4,4} \\
Y^{(2)}_{\mathbf{6}_4,5} & -\sqrt{2}\,Y^{(2)}_{\mathbf{6}_4,1} & -Y^{(2)}_{\mathbf{6}_4,6} \\
-Y^{(2)}_{\mathbf{6}_4,6} & -\sqrt{2}\,Y^{(2)}_{\mathbf{6}_4,2} & Y^{(2)}_{\mathbf{6}_4,5}
\end{pmatrix}
\, \sim \,
\begin{pmatrix}
2\sqrt{2}\,\epsilon^8 & 0 & -\sqrt{2} \\
4\sqrt{2}\,\epsilon^5 & -20\,\epsilon^{11} & 12\sqrt{2}\,\epsilon^{13} \\
12\sqrt{2}\,\epsilon^{13} & -4\,\epsilon^3 & 4\sqrt{2}\,\epsilon^5
\end{pmatrix}
\,+\, \mathcal{O}(\epsilon^{16})
\,,
\label{eq:Md}
\end{align}
in a right-left convention. Both determinants are non-vanishing and seen to be $\mathcal{O}(\epsilon^8)$. As anticipated, taking into account the precise structure of the mass matrices, one obtains
\begin{equation} \label{eq:H2H5}
\begin{aligned}
m_e:m_\mu:m_\tau \;&=\; 4\sqrt{2}\,\epsilon^{6} : 2\,\epsilon^{2} : 1\,,
\quad\qquad (\text{H}_2)
\\
m_d:m_s:m_b \;&=\; 4\,\epsilon^{5} : 2\sqrt{2}\,\epsilon^{3} : 1\,,
\quad\qquad (\text{H}_5)
\end{aligned}
\end{equation}
at leading order. Corrections to the spectra, including a possible dependence on $\re \tau$, emerge only at $\mathcal{O}(\epsilon^{12})$ and $\mathcal{O}(\epsilon^{10})$, respectively. For a given value of $\epsilon = e^{-\pi \im \tau / 8}$, the mass hierarchies are fixed \emph{exactly} by the modular structure, i.e.~without invoking additional $\mathcal{O}(1)$ flavour coefficients beyond the single, overall parameters $\alpha_{d,e}$.

\vskip 2mm

The charged-lepton and down-quark mass matrices are controlled by only three continuous parameters: two overall normalisations, which can be fixed e.g.\ by $m_\tau$ and $m_b$, and the common expansion parameter $\epsilon$. Therefore, the resulting hierarchies lead to three exact mass relations at the flavour-breaking scale $M_F\gg \Lambda_{\rm EW}$. It is convenient to start from the two \emph{intra-sector} relations, which are found separately, within each sector:
\begin{align}
R_q:\quad
1 &=
\frac{1}{2\sqrt{2}}\,
\frac{m_s^5}{m_b^2\,m_d^3}\,,
\label{eq:mass_relation_down}
\\[2mm]
R_\ell:\quad
1 &=
\frac{1}{\sqrt{2}}\,
\frac{m_\mu^3}{m_\tau^2\,m_e}\,.
\label{eq:mass_relation_CL}
\end{align}
These relations can be rewritten as $m_s^5 = 2\sqrt{2}\,m_d^3m_b^2$ and $m_\mu^3 = \sqrt{2}\,m_e m_\tau^2$, and follow directly from~\cref{eq:H2H5}. One may additionally form \emph{inter-sector} relations. Indeed, the fact that the same $\epsilon$ controls both sectors implies
\begin{align}
R_{\ell q}:\quad
1 &=
\frac{1}{\sqrt{2}}\,
\frac{m_s^2}{m_b^2}\,
\frac{m_\tau}{m_e}\,,
\label{eq:mass_relation_inter_e}
\end{align}
that can be rewritten as $m_s^2m_\tau = \sqrt{2}\,m_e m_b^2$, while other such relations can be obtained algebraically. In particular, combining~\cref{eq:mass_relation_CL,eq:mass_relation_down} yields a relation involving all six masses,
\begin{align}
\frac{m_e\, m_\tau^2}{m_\mu^3} \;=\; 2\,\frac{m_d^3\, m_b^2}{m_s^5}\,,
\label{eq:mass_relation_all6}
\end{align}
and one can also find
\begin{align}
\frac{m_\mu}{m_\tau} \;=\; \sqrt{2}\,\frac{m_d}{m_s}\,.
\label{eq:mass_relation_inter_mu}
\end{align}

An explicit non-minimal completion of the up-quark sector is given in Appendix~\ref{app:up-completion}. Keeping the same $Q$ representation and modulus as in the down sector, a weight-10 up-quark Yukawa matrix spans an eight-dimensional subspace of complex $3\times3$ matrices. We show that this subspace contains a matrix with the observed up-quark masses and CKM parameters. These observables are fitted by complex coefficients, while the down-quark and charged-lepton \ac{OPM} mass relations remain unchanged.

%%%%%%%%%%%%%%%%%%%%%%%%%%%%%%%%
\section{Running and threshold corrections}
\label{sec:phenomenology}
%%%%%%%%%%%%%%%%%%%%%%%%%%%%%%%%

The mass relations derived above are exact at the flavour scale $M_F$, where the modular symmetry is imposed in the supersymmetric theory. At this scale, the charged-lepton and down-quark Yukawa structures are fixed in terms of a single parameter $\epsilon$. To compare these predictions with low-energy data, one must account for the \ac{RG} evolution between $M_F$ and the electroweak scale. We take $M_F=2\times 10^{16}\,\mathrm{GeV}$ and consider a supersymmetric completion in which the \ac{RG} evolution is governed by MSSM beta functions above the intermediate scale $M_{\rm SUSY}$, and by SM beta functions below it. We assume that possible direct SUSY-breaking corrections to the high-scale modular Yukawa operators are negligible~\cite{Criado:2018thu}, so that the \ac{OPM} relations define the boundary conditions for the dimensionless Yukawa couplings. We do not specify the mediation mechanism of SUSY breaking. Instead, we first isolate the effect of \ac{RG} running, and later include finite SUSY threshold corrections at $M_{\rm SUSY}$, treating the soft spectrum phenomenologically.

%%%%%%%%%%%%%%%%%%%%%%%%%%%%%%%%%%%%%%%%%%
\subsection{Running without finite threshold corrections}
\label{subsec:running_no_tc}
%%%%%%%%%%%%%%%%%%%%%%%%%%%%%%%%%%%%%%%%%%

We first neglect finite SUSY threshold corrections. This provides a clean diagnostic of which part of the flavour pattern is already captured by the one-parameter structure before invoking matching effects.

\begin{figure}[t]
\centering
\begin{subfigure}{0.48\textwidth}
\centering
\includegraphics[width=\linewidth]{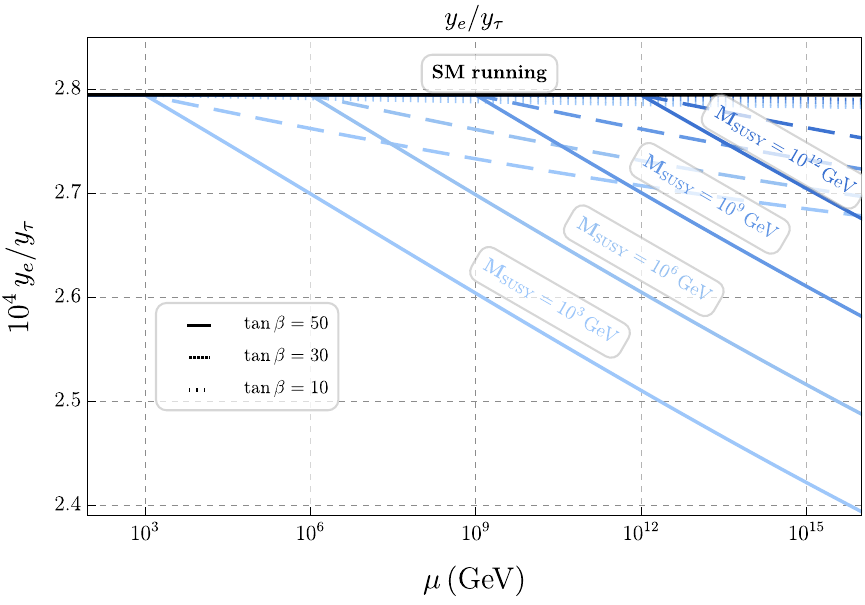}
\caption{}
\label{fig:running-exp-ye}
\end{subfigure}
\hfill
\begin{subfigure}{0.48\textwidth}
\centering
\includegraphics[width=\linewidth]{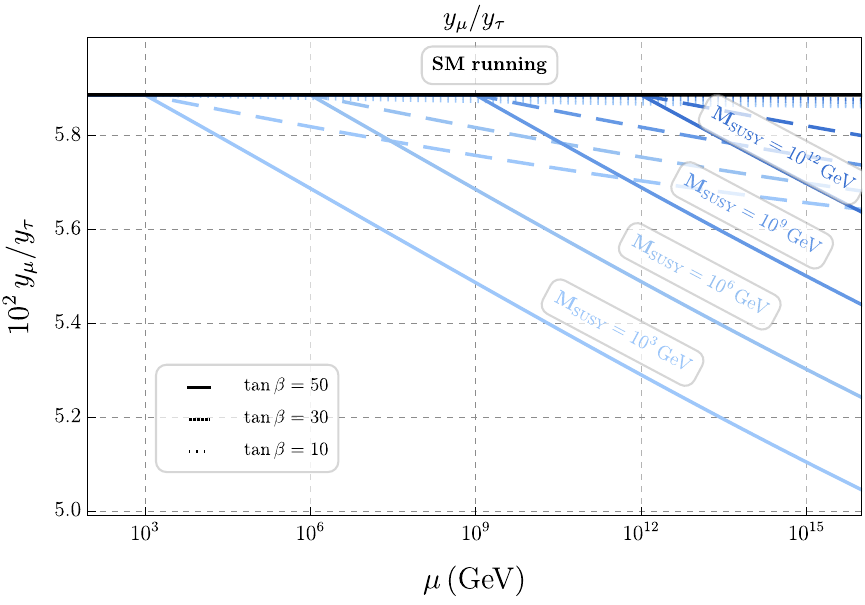}
\caption{}
\label{fig:running-exp-ymu}
\end{subfigure}

\vspace{0.3cm}

\begin{subfigure}{0.48\textwidth}
\centering
\includegraphics[width=\linewidth]{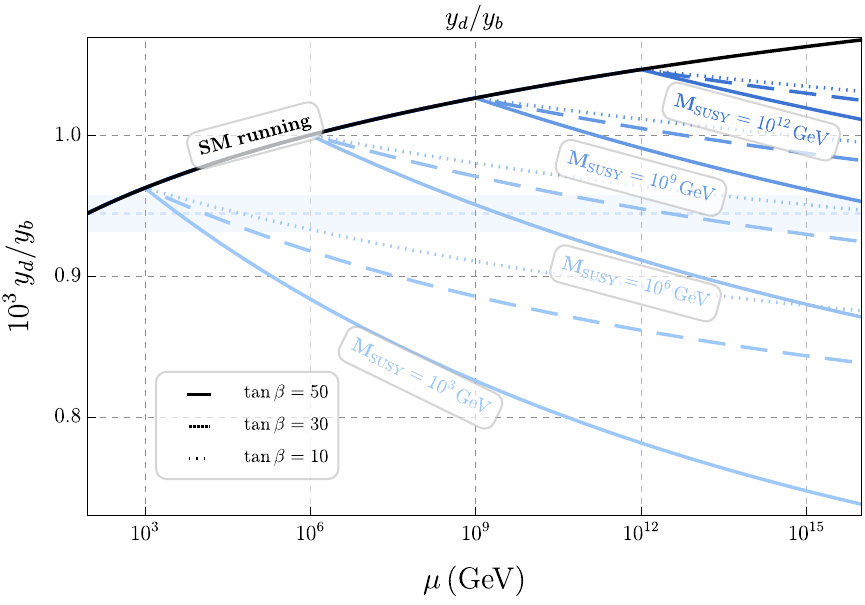}
\caption{}
\label{fig:running-exp-yd}
\end{subfigure}
\hfill
\begin{subfigure}{0.48\textwidth}
\centering
\includegraphics[width=\linewidth]{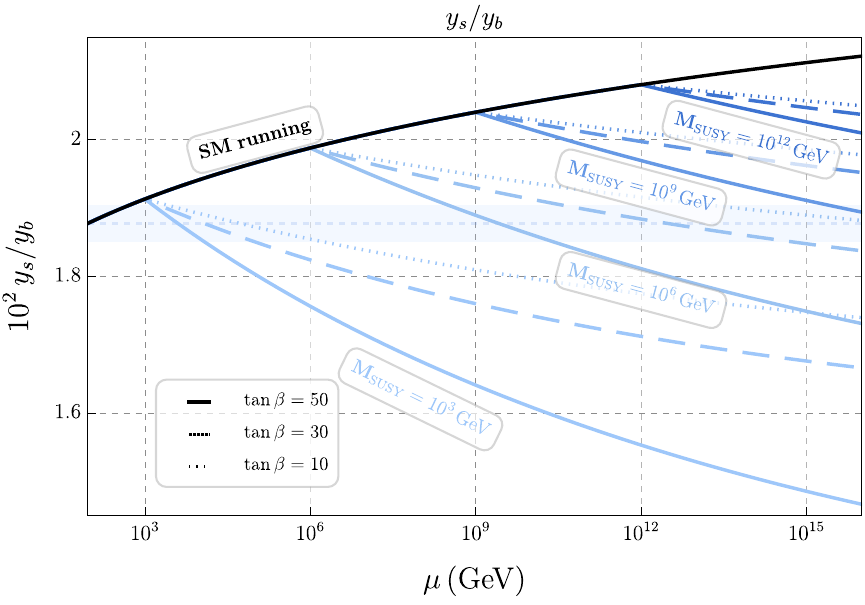}
\caption{}
\label{fig:running-exp-ys}
\end{subfigure}

\caption{ \ac{RG} evolution of the experimentally extracted Yukawa ratios from $M_Z$ to $M_F=2\times10^{16}\,\mathrm{GeV}$, for different choices of $M_{\rm SUSY}$ and $\tan\beta$. The evolution is performed with SM \acp{RGE} below $M_{\rm SUSY}$ and MSSM \acp{RGE} above $M_{\rm SUSY}$, without finite SUSY threshold corrections. }
\label{fig:pedagogical-running}
\end{figure}

We begin with a purely pedagogical exercise. Starting from the experimentally extracted Yukawa ratios at $M_Z$, we evolve them upwards to $M_F=2\times 10^{16}\,\mathrm{GeV}$, using SM \acp{RGE} below $M_{\rm SUSY}$ and MSSM \acp{RGE} above $M_{\rm SUSY}$. We use \texttt{REAP}~\cite{Antusch:2005gp} and \texttt{SUSYTC}~\cite{Antusch:2015nwi,Antusch:2020ztu} to compute the running. The result is shown in~\Cref{fig:pedagogical-running} for different choices of $M_{\rm SUSY}$ and $\tan\beta$. The running of the ratios is clearly sector-dependent: the charged-lepton ratios are almost scale-independent in the SM regime, but can receive visible MSSM running above $M_{\rm SUSY}$, especially at large $\tan\beta$; the down-quark ratios also show a significant dependence on both $M_{\rm SUSY}$ and $\tan\beta$. Thus the comparison between the high-scale modular relations and low-energy data is not SUSY-scale-independent. The parameters $M_{\rm SUSY}$ and $\tan\beta$ affect the image of the low-energy data at $M_F$, and hence the value of $\epsilon$ selected by the observed hierarchies.

\begin{figure}[t]
\centering
\includegraphics[width=0.58\textwidth]{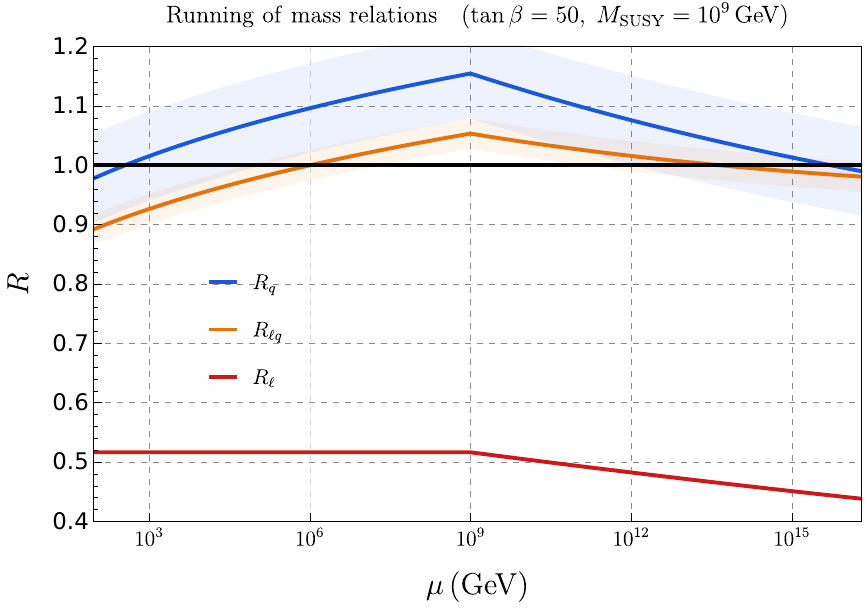}
\caption{Running of the mass relations in~\cref{eq:mass_relation_down,eq:mass_relation_inter_e,eq:mass_relation_CL}. Recall that the \ac{OPM} predicts $R_q=R_{\ell q}=R_\ell=1$ at the flavour scale $M_F$. The experimental Yukawa ratios and their uncertainties (shaded bands) are extracted at $M_Z$ and evolved upwards to $M_F=2\times 10^{16}\,{\rm GeV}$, using SM \acp{RGE} below $M_{\rm SUSY}$ and MSSM \acp{RGE} above $M_{\rm SUSY}$, without finite threshold corrections. The plot is shown for a representative choice of $M_{\rm SUSY}$ and $\tan\beta$. The down-sector and inter-sector relations are close to unity within the running uncertainties, while the charged-lepton relation, controlled by the muon-to-tau hierarchy, remains clearly displaced.}
\label{fig:mass-relations-running}
\end{figure}

We then run the mass relations of~\cref{eq:mass_relation_down,eq:mass_relation_CL,eq:mass_relation_inter_e} from $M_Z$ to $M_F$, for a particular choice of $M_{\rm SUSY} = 10^9$ GeV and $\tan\beta=50$. The result is shown in~\Cref{fig:mass-relations-running}. In this plot, no model prediction is imposed yet as a boundary condition at $M_F$. Instead, it asks whether the low-energy data, when lifted to high scales, can approach the relations predicted at the flavour scale. The result already captures the central pattern. The down-sector relation of~\cref{eq:mass_relation_down} is compatible with unity within the propagated uncertainty over a broad range of SUSY scales and values of $\tan \beta$, while the inter-sector relation of~\cref{eq:mass_relation_inter_e} is also close to unity for $M_{\rm SUSY}\sim 10^9$ GeV and high $\tan \beta$. By contrast, the charged-lepton relation of~\cref{eq:mass_relation_CL} remains far from unity and is not rescued by running. This identifies the muon mass prediction as the main discrepancy: the model tends to predict a muon mass around $20\%$ higher than the experimentally measured one.

We now impose the high-scale one-parameter predictions explicitly. At $M_F$, the relevant ratios are
\begin{align}
    \left.\frac{y_d}{y_b}\right|_{M_F}=4\,\epsilon^5\,,
    \qquad
    \left.\frac{y_s}{y_b}\right|_{M_F}=2\sqrt{2}\,\epsilon^3\,,
    \label{eq:down_ratios_MF}\\
    \left.\frac{y_e}{y_\tau}\right|_{M_F}=4\sqrt{2}\,\epsilon^6\,,
    \qquad
    \left.\frac{y_\mu}{y_\tau}\right|_{M_F}=2\,\epsilon^2\,.
    \label{eq:lepton_ratios_MF}
\end{align}
For each pair $(\epsilon,M_{\rm SUSY})$, we impose these relations at $M_F$, run the Yukawa matrices down to $M_Z$, and compare with the experimental values at $M_Z$. The behaviour of the mass relations in \Cref{fig:mass-relations-running} motivates a first fit in which the muon ratio is not included. We therefore fit only
\begin{equation}
    \frac{y_d}{y_b}\,,\qquad
    \frac{y_s}{y_b}\,,\qquad
    \frac{y_e}{y_\tau}\,,
\end{equation}
and leave $y_\mu/y_\tau$ as an independent diagnosis of the threshold corrections that will be needed. Since charged-lepton SUSY thresholds are $\tan \beta$-enhanced, we use $\tan \beta = 50$ as a benchmark for the threshold analysis. The best-fit point is
\begin{equation}
    \epsilon_{\rm best}=0.1892\,,    \footnote{The best-fit value $\epsilon_{\rm best}=0.1892$ corresponds to ${\rm Im}\tau\simeq 4.24$ for $N=16$. Thus the expansion in $q$ is very well controlled, $|q|\simeq 2.7\times 10^{-12}$, while the modulus is only moderately close to the cusp. We therefore do not rely on an extreme decompactification regime, although the UV interpretation of this region is model-dependent.}
    \qquad
    M_{\rm SUSY}^{\rm best}\simeq 5.8\times 10^9~\mathrm{GeV}\,.
    \label{eq:best_fit_no_tc}
\end{equation}

\begin{figure}[t]
\centering
\begin{subfigure}{0.48\textwidth}
\centering
\includegraphics[width=\linewidth]{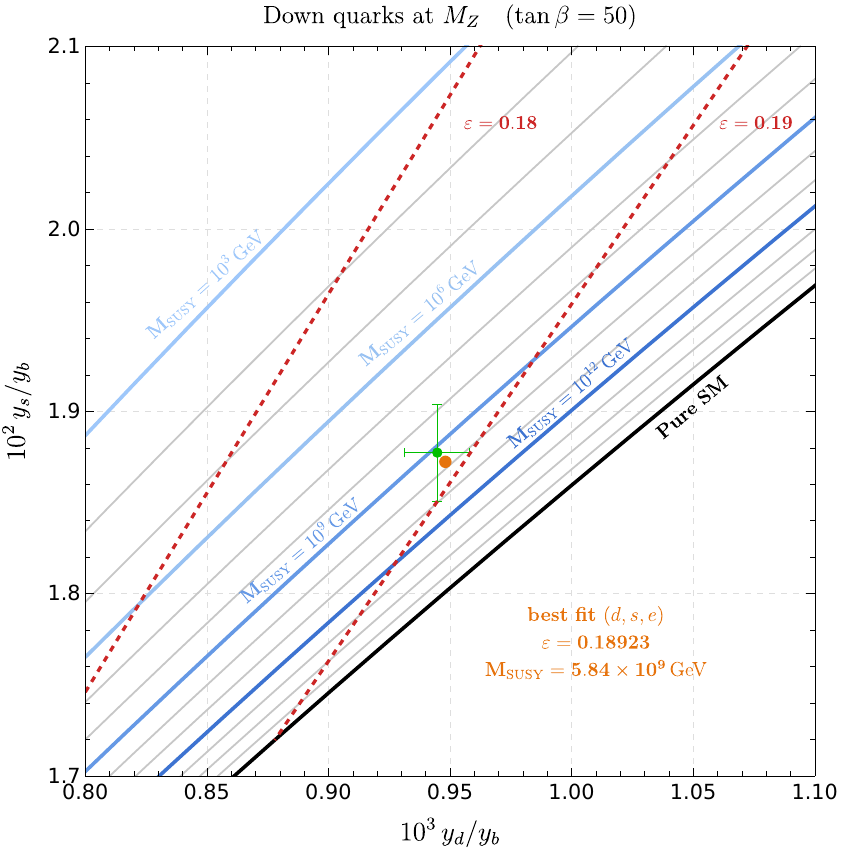}
\caption{}
\label{fig:epsilon-down}
\end{subfigure}
\hfill
\begin{subfigure}{0.48\textwidth}
\centering
\includegraphics[width=\linewidth]{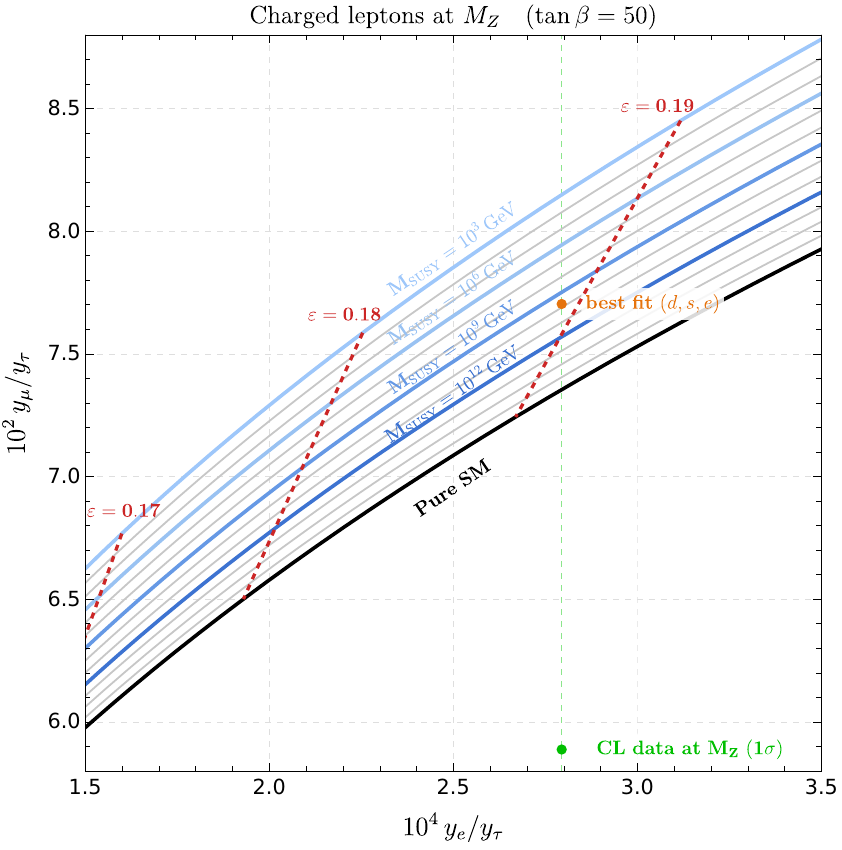}
\caption{}
\label{fig:epsilon-leptons}
\end{subfigure}
\caption{ Low-energy predictions of the \ac{OPM} before finite SUSY threshold corrections. The curves show the image at $M_Z$ of the high-scale relations as $\epsilon$ and $M_{\rm SUSY}$ are varied. The best-fit point is obtained by fitting $y_d/y_b$, $y_s/y_b$, and $y_e/y_\tau$, leaving $y_\mu/y_\tau$ as a prediction. The result is $\epsilon \simeq 0.189$ and $M_{\rm SUSY} \simeq 10^{9.77}$ GeV. The fitted ratios are brought into agreement with data, while the muon-to-tau ratio remains significantly displaced, identifying the required threshold correction.   }
\label{fig:epsilon-predictions}
\end{figure}

The result of the fit is shown in \Cref{fig:epsilon-predictions}. With only two continuous parameters, $\epsilon$ and $M_{\rm SUSY}$, the model brings the two down-quark ratios and $y_e/y_\tau$ within their corresponding $1\sigma$ experimental regions. This is already a non-trivial alignment between the down-quark and charged-lepton sectors. The same point, however, predicts $y_\mu/y_\tau$ significantly above the experimental value. Thus the no-threshold analysis isolates a single problem: the model captures the down-quark hierarchy and the electron-to-tau hierarchy, while the muon-to-tau ratio requires an additional effect that may arise from matching.

%%%%%%%%%%%%%%%%%%%%%%%%%%%%%%%%%%%%%%%%%%
\subsection{Targeted charged-lepton threshold correction}
\label{subsec:muon_tc}
%%%%%%%%%%%%%%%%%%%%%%%%%%%%%%%%%%%%%%%%%%

The no-threshold result shows that we do not need arbitrary threshold effects to rescue the full mass pattern: the double \ac{OPM} structure already captures the two down-quark ratios and the electron-to-tau ratio with a common $\epsilon$. The remaining discrepancy is isolated in the muon-to-tau ratio. In the following we adopt a bottom-up treatment of the soft supersymmetry-breaking sector. The soft terms are taken to be generic matching parameters at $M_{\rm SUSY}$, and are not assumed to follow from a modular-covariant supersymmetry-breaking sector. They therefore act as additional sources of modular-flavour breaking. A more constrained possibility would be to impose  modular covariance also on the soft terms, see e.g.~\cite{Kikuchi:2022pkd,Ding:2022nzn}. We leave this direction for future work.

We parametrize the finite matching correction to charged-lepton Yukawa ratios at $M_{\rm SUSY}$ as
\begin{equation}
    \left.\frac{y_{\ell_i}}{y_\tau}\right|_{\mathrm{SM}}
    =
    \kappa_i
    \left.\frac{y_{\ell_i}}{y_\tau}\right|_{\mathrm{MSSM}}\,,
    \qquad
    \kappa_i\equiv \frac{1+\Delta_i}{1+\Delta_\tau}
    \qquad\quad (i=e,\mu)\,.
    \label{eq:kappa_def}
\end{equation}
The best-fit point described above requires
\begin{equation}
    \kappa_e=1\,,
    \qquad
    \kappa_\mu=0.764\,,
    \label{eq:kappa_required}
\end{equation}
up to the quoted numerical precision. Thus the required correction preserves $y_e/y_\tau$ and modifies only $y_\mu/y_\tau$ relative to $y_\tau$. In the quark sector we assume flavour-universal threshold corrections, so that the down-quark ratios $y_d/y_b$ and $y_s/y_b$ are also preserved.

We now show that a correction of this form can be generated by standard one-loop electroweak SUSY threshold corrections. Following Ref.~\cite{Antusch:2008tf}, and working in the large-$\tan\beta$ regime while neglecting trilinear terms, the charged-lepton correction can be written as
\begin{equation}
    \Delta_i
    =
    \frac{\tan\beta}{16\pi^2}
    \left[
        \tau_i^W+\tau_i^B
    \right]\,,
    \label{eq:Delta_l_def}
\end{equation}
with
\begin{equation}
    \tau_i^W
    =
    \frac{3}{2} g_2^2
    \frac{M_2}{\mu_H}
    h_2\!\left(
        \frac{M_2^2}{\mu_H^2},
        \frac{m_{L_i}^2}{\mu_H^2}
    \right)\,,
    \label{eq:tauW}
\end{equation}
and
\begin{equation} \label{eq:tauB}
    \tau_i^B
    =
    \frac{3}{5}g_1^2
    \bigg[
    -\frac{\mu_H}{M_1}
    h_2\!\left(
        \frac{m_{e_i}^2}{M_1^2},
        \frac{m_{L_i}^2}{M_1^2}
    \right)
    +
    \frac{M_1}{\mu_H}
    h_2\!\left(
        \frac{m_{e_i}^2}{\mu_H^2},
        \frac{M_1^2}{\mu_H^2}
    \right)
    -
    \frac{1}{2}
    \frac{M_1}{\mu_H}
    h_2\!\left(
        \frac{M_1^2}{\mu_H^2},
        \frac{m_{L_i}^2}{\mu_H^2}
    \right)
    \bigg]\,.
\end{equation}
Here $M_1$ and $M_2$ are the bino and wino masses, $\mu_H$ is the higgsino mass parameter, and $m_{L_i}$, $m_{e_i}$ are the left- and right-handed charged-slepton soft masses. The loop function is
\begin{equation}
    h_2(x,y)
    \equiv
    \frac{x\log x}{(1-x)(x-y)}
    +
    \frac{y\log y}{(1-y)(y-x)}\,,
    \label{eq:h2_loop}
\end{equation}
with the degenerate limits understood by continuity.

For our present purpose it is sufficient to consider a simple analytic ansatz. We take
\begin{equation}
    M_1=M_2=\mu_H\equiv M\,,
    \label{eq:tc_ansatz_gauginos}
\end{equation}
and
\begin{equation}
    m_{e_1}=m_{L_1}=m_{e_3}=m_{L_3}
    \equiv x_{e\tau}M\,,
    \qquad
    m_{e_2}=m_{L_2}
    \equiv x_\mu M\,.
    \label{eq:tc_ansatz_sleptons}
\end{equation}
This ansatz makes the first and third charged-lepton generations identical from the point of view of threshold corrections. Therefore $\Delta_e=\Delta_\tau$ and $\kappa_e=1$ as required. The muon correction is instead controlled by the single ratio $x_\mu$, while the common electron--tau threshold is controlled by $x_{e\tau}$.

Under~\cref{eq:tc_ansatz_gauginos,eq:tc_ansatz_sleptons}, the correction becomes a function only of the dimensionless ratio $x_i=m_i/M$,
\begin{equation}
    \Delta(x_i)
    \,=\,
    \frac{\tan\beta}{16\pi^2}
    \bigg\{
    \frac{3}{2}g_2^2 h_2(1,x_i^2)
    +
    \frac{3}{5}g_1^2
    \left[
        -h_2(x_i^2,x_i^2)
        +h_2(x_i^2,1)
        -\frac{1}{2}h_2(1,x_i^2)
    \right]
    \bigg\}\,.
    \label{eq:Delta_x}
\end{equation}
The relevant matching factors are then
\begin{equation}
    \kappa_e=1\,,
    \qquad
    \kappa_\mu
    =
    \frac{1+\Delta(x_\mu)}
         {1+\Delta(x_{e\tau})}\,.
    \label{eq:kappamu_x}
\end{equation}
Using the gauge couplings evaluated at the best-fit matching scale,
\begin{equation}
    g_1(M_{\rm SUSY}^{\rm best})=0.515\,,
    \qquad
    g_2(M_{\rm SUSY}^{\rm best})=0.5703\,,
    \qquad
    \tan\beta=50\,,
    \label{eq:gauge_values_tc}
\end{equation}
we find that the simple choice
\begin{equation}
    x_\mu=0.5\,,
    \qquad
    x_{e\tau}=0.012137\,,
    \label{eq:tc_benchmark_x}
\end{equation}
gives
\begin{equation}
    \frac{1+\Delta_\mu}{1+\Delta_\tau}
    \simeq
    0.764\,,
    \qquad
    \frac{1+\Delta_e}{1+\Delta_\tau}=1\,.
    \label{eq:tc_benchmark_result}
\end{equation}
For both left- and right-handed sleptons, the values in \eqref{eq:tc_benchmark_x} correspond to the soft-mass hierarchy
\[
\frac{m_{\widetilde{\mu}_{L,R}}}{m_{\widetilde{e}_{L,R}}}=\frac{m_{\widetilde{\mu}_{L,R}}}{m_{\widetilde{\tau}_{L,R}}}=\frac{x_\mu}{x_{e\tau}}\simeq 41.
\]
The illustrative benchmark is therefore strongly hierarchical. This hierarchy is not a prediction of the \ac{OPM} construction, but a property of this particular realization; different soft spectra may produce the same effective matching correction. Accordingly, this benchmark should be viewed as an explicit proof of existence. A fully specified realization of such a split spectrum would require the corresponding multi-scale matching treatment, which lies beyond the scope of the present work.

In the down sector we take the squark soft masses to be flavour universal, $m_{Q_i}=m_Q$ and $m_{d_i}=m_d$. The corresponding finite threshold corrections are then generation independent, $\Delta_d=\Delta_s=\Delta_b$, so that they cancel in the ratios $y_d/y_b$ and $y_s/y_b$. The down-quark ratios are protected by construction, while the only non-universal correction relevant for our purpose is the muonic one in~\cref{eq:tc_benchmark_result}. This provides a proof of existence: a selective muon threshold correction of the required size can be generated at one loop, while preserving the successful electron-to-tau and down-quark ratios.

\begin{figure}[p]
\vbox to \textheight{ \vfill \includegraphics[width=0.78\textwidth]{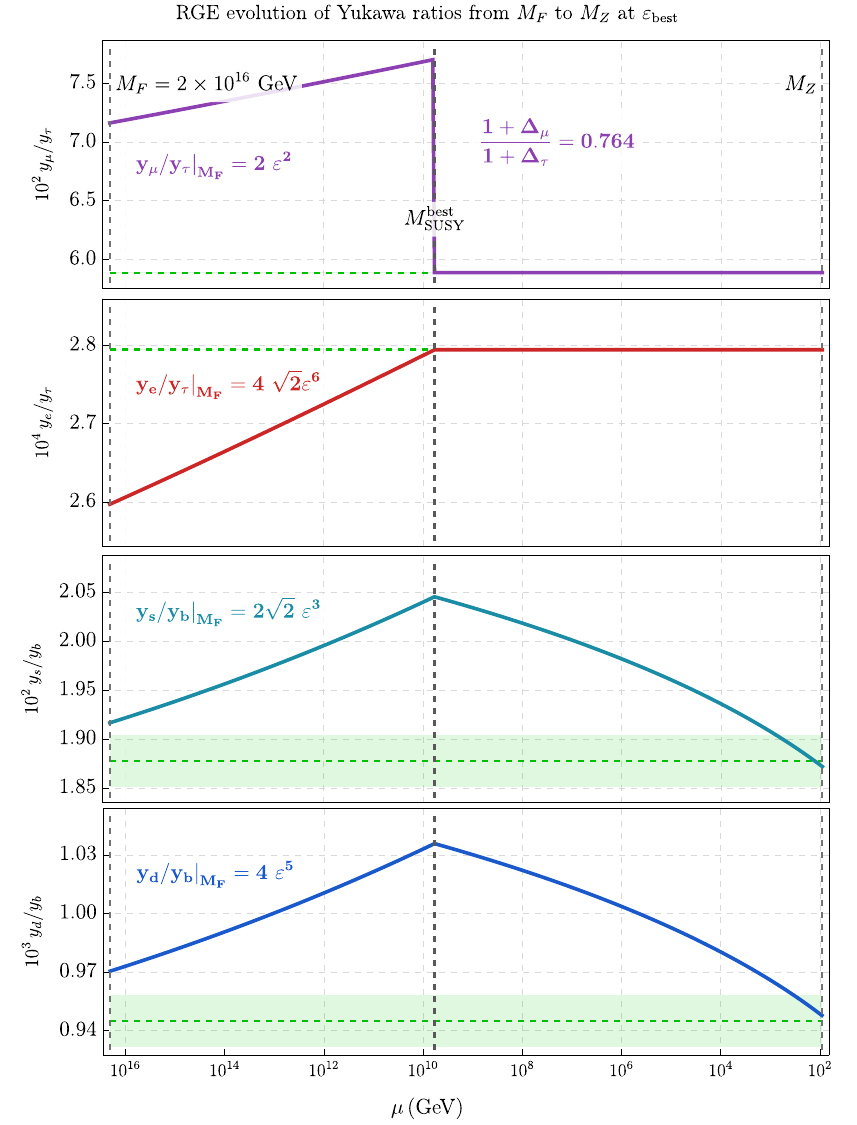} \caption{\ac{RG} evolution of the rescaled Yukawa ratios from $M_F=2\times10^{16}\,\mathrm{GeV}$ to $M_Z$ at $\epsilon_{\rm best}$ and $M_{\rm SUSY}^{\rm best}$. The green bands denote the experimental $1\sigma$ intervals at $M_Z$. The high-scale relations for $y_d/y_b$, $y_s/y_b$, and $y_e/y_\tau$ evolve into the corresponding experimental bands, while the muon ratio requires a selective threshold correction. The displayed benchmark applies $(1+\Delta_\mu)/(1+\Delta_\tau)=0.764$ at $M_{\rm SUSY}^{\rm best}$.} \label{fig:final-running-with-muon-tc} \vfill }
\end{figure}

The complete running of the flavour prediction from $M_F$ to $M_Z$ including this effective threshold correction is shown in \Cref{fig:final-running-with-muon-tc}. At $M_F$, the four ratios are fixed by the one-parameter modular relations. They are then evolved down to $M_{\rm SUSY}^{\rm best}$, where the matching factor in~\cref{eq:tc_benchmark_result} is applied to the muon-to-tau ratio. After subsequent SM running down to $M_Z$, all four ratios lie in the corresponding experimental bands. The need for sizeable threshold corrections shifts the prediction of the model from a sharp flavour prediction to a constraint on the SUSY spectrum.

%%%%%%%%%%%%%%%%%%%%%%%%%%%%%
\section{A possible one-parameter up-quark hierarchy}
\label{sec:quark-flavour-hints}
%%%%%%%%%%%%%%%%%%%%%%%%%%%%%

The charged-lepton and down-quark \ac{OPM} sectors can be combined with the explicit up-quark completion of Appendix~\ref{app:up-completion}. In that construction, the up-quark masses and CKM observables are fitted by complex coefficients. We examine here the separate possibility that the up-quark hierarchy might itself arise from an \ac{OPM}.

A first important point is that the one-parameter predictions are genuinely restrictive. At the flavour scale $M_F$, each \ac{OPM} hierarchy defines a one-dimensional curve in the hierarchy plane
\[
\left(
\frac{m_1}{m_3},
\frac{m_2}{m_3}
\right).
\]
The comparison with data is not completely direct, since the flavour relations are imposed at $M_F$, whereas the charged-fermion masses are measured at low energies. In a supersymmetric setup, this introduces several additional ingredients: the flavour scale $M_F$, the SUSY scale $M_{\rm SUSY}$, $\tan\beta$, and possible finite threshold corrections at $M_{\rm SUSY}$. One may therefore worry that these effects could wash out the one-parameter predictions. However, we should emphasize that the role of $\epsilon$ is qualitatively different from that of the parameters entering RG evolution and threshold matching. While $\epsilon$ fixes the characteristic powers and correlations defining each \ac{OPM} curve, the latter affect only its precise mapping to low energies. Their relevance for the comparison with data reflects the predictivity of the \ac{OPM} framework, rather than an additional freedom to generate the observed hierarchies.

\begin{figure}[p]
\vbox to \textheight{ \vfill \includegraphics[width=0.72\textwidth]{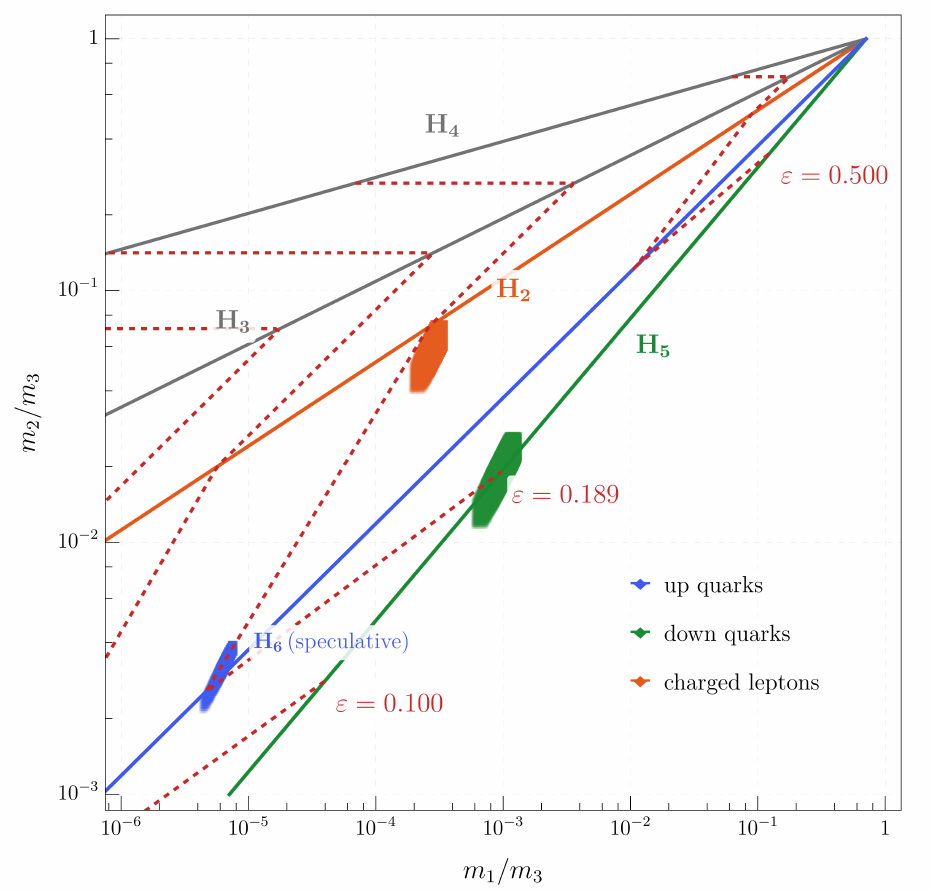} \caption{ High-scale hierarchy plane illustrating the non-triviality of the one-parameter predictions and the possible up-quark hint. The coloured regions are obtained by lifting the measured low-energy up-quark, down-quark and charged-lepton hierarchies taken from \cite{ParticleDataGroup:2024cfk} to the high-scale plane, scanning over $M_F$, $M_{\rm SUSY}$, and $\tan\beta$, and applying threshold-like deformations to the individual Yukawa eigenvalues. We use $\Delta_i\in[-0.15,0.15]$ for the down-quark and charged-lepton sectors, and $\Delta_i^u\in[-0.05,0.05]$ for the up-quark sector. Even with this enlarged freedom, the allowed regions remain small compared with the full hierarchy plane. The found \ac{OPM} curves show that H$_5$ naturally matches the down-quark hierarchy, while H$_2$ lies close to the charged-lepton region once sizeable muon-threshold effects are allowed. The hypothetical H$_6$ curve, with $p/q=2$, passes close to the up-quark region for the same order of $\epsilon$, namely $\epsilon\sim\theta_C$. } \label{fig:hierarchy-blobs} \vfill }
\end{figure}

\afterpage{\clearpage}

We test explicitly whether these effects can wash out the one-parameter predictions in \Cref{fig:hierarchy-blobs} and find that they do not. Instead of running each model curve down to $M_Z$, we lift the measured low-energy hierarchies up to the high-scale hierarchy plane. For each charged-fermion sector ($f=u,d,e$) we start from
\[
\left(
\frac{m_1^f}{m_3^f},
\frac{m_2^f}{m_3^f}
\right)_{M_Z}\,,
\]
and evolve the ratios upward, scanning over representative values of $M_F$, $M_{\rm SUSY}$, and $\tan\beta$. We also include a deliberately aggressive threshold broadening. For the down-quark and charged-lepton sectors, we allow
\[
\Delta_i^e,\,\Delta_i^d\,\in\,[-0.15,0.15]\,,
\]
while for the up-quark sector we use the smaller range
\[
\Delta_i^u\,\in\,[-0.05,0.05]\,,
\]
since these corrections are not $\tan\beta$-enhanced. At the level of ratios, these deformations act as
\[
\frac{m_1}{m_3}
\,\longrightarrow\,
\frac{m_1}{m_3}
\frac{1+\Delta_3}{1+\Delta_1}\,,
\qquad
\frac{m_2}{m_3}
\,\longrightarrow\,
\frac{m_2}{m_3}
\frac{1+\Delta_3}{1+\Delta_2}\,.
\]
This prescription is not meant to represent a detailed threshold calculation, but as a stress test: we intentionally enlarge the regions selected by the data in order to see whether a generic one-parameter curve could be made viable by running and threshold freedom alone.

The allowed regions remain small even after this deliberately generous broadening. In other words: running effects, the choice of $M_F$, $M_{\rm SUSY}$, $\tan\beta$, and sizeable threshold-like deformations do not turn the hierarchy plane into a free fit. Most one-parameter curves still miss the data. The cases that do work therefore represent non-trivial alignments between the modular prediction and the charged-fermion hierarchies.

The hierarchy patterns found in~\Cref{sec:models} provide a further clue. As discussed around~\cref{eq:meta}, the \ac{OPM} hierarchies obtained so far are not arbitrary curves in the hierarchy plane, but follow a common meta-relation given by~\cref{eq:meta}. In the hierarchy plane, the ratio $p/q$ controls the slope of the corresponding one-parameter curve. The down-quark hierarchy is naturally matched by H$_5$, while the charged-lepton hierarchy is close to H$_2$ once a muon threshold correction is included. Interestingly, the up-quark hierarchy lies close to the direction selected by
\[
\frac{p}{q}=2\,.
\]

This motivates the following hypothetical hierarchy, obtained by taking $p/q=2$ and imposing the relation in~\cref{eq:meta},
\begin{equation}
\text{H}_6:\qquad
m_u:m_c:m_t
=
4\sqrt{2}\,\epsilon^8:
2\sqrt{2}\,\epsilon^4:
1\,.
\end{equation}
This pattern is not one of the \ac{OPM} hierarchies explicitly constructed in~\Cref{sec:models}. Nevertheless, it is suggestive for two reasons. First, it follows the simple (integer) $p/q=2$ scaling selected by the up-quark hierarchy. Second, it intersects the up-quark region for the same value of $\epsilon$ as the one selected by the charged-lepton and down-quark sectors. Numerically, this common value is of the order of the Cabibbo angle,
\[
\epsilon \sim \theta_C\,.
\]
Taken together, these observations suggest a possible route towards the quark flavour puzzle. One may imagine a modular construction in which the down-quark sector is controlled by the H$_5$ hierarchy, the up-quark sector by an H$_6$-type hierarchy, and the CKM structure arises from mixing effects controlled by the same small parameter $\epsilon \sim \theta_C$. We stress that this is a speculative observation. We have not constructed an explicit model realizing H$_6$ and do not know whether such a model can also reproduce quark mixing and the corresponding CP violation phase.\footnote{In this restrictive context, ($\re\tau$)-dependent terms in the $q$-expansions may not be enough to generate sufficient CP violation, calling instead for complex \ac{CG} coefficients that signal CP violation at the group level. This would introduce an additional selection criterion on candidate finite modular groups.} In such a hypothetical scenario, the observed quark mass hierarchies and the size of quark mixing would have a common modular origin. In addition to the down-quark mass relation of~\cref{eq:mass_relation_down} we would have
\begin{equation}
m_c^2 = \sqrt{2} m_u \, m_t\,, \qquad  \frac{m_s^4}{m_b^4}  = 2 \sqrt{2} \, \frac{m_c^3}{m_t^3}\,.
\label{eq:mass_relation_upandupdown}
\end{equation}

Moreover, if the charged leptons are also assigned to the H$_2$ hierarchy, the $5$ mass relations of~\cref{eq:mass_relation_down,eq:mass_relation_CL,eq:mass_relation_inter_e,eq:mass_relation_upandupdown} would be satisfied at the same flavour scale. Alternatively, by algebraically combining those we could build a mass relation involving \textit{all} the $9$ charged fermions,
\begin{equation}
\frac{1}{2\sqrt{2}} \left(\frac{m_e m_\mu}{m_\tau^2}\right)
\left(\frac{m_d m_s}{m_b^2}\right)
\left(\frac{m_t^3}{m_u m_c^2}\right)
=
1\,.
\end{equation}
Using the PDG values and $1\sigma$ uncertainties at the $M_Z$ scale~\cite{ParticleDataGroup:2024cfk}, we find
\begin{equation}
\frac{1}{2\sqrt{2}} \left.\left(\frac{m_e m_\mu}{m_\tau^2}\right)
\left(\frac{m_d m_s}{m_b^2}\right)
\left(\frac{m_t^3}{m_u m_c^2}\right)\right|_{M_Z}
=1.05 \pm 0.05\,.
\label{eq:massrels9MZ}
\end{equation}

However, at present this remains speculative as we have not found an explicit \ac{OPM} realizing the H$_6$ hierarchy. Finally, while~\cref{eq:meta} holds for the models based on the finite modular groups $\Delta(96)$ and $\Delta(384)$, it is not guaranteed to apply to all conceivable \acp{OPM}.

%%%%%%%%%%%%%%%%%%%%%%%%%%%%%
\section{Conclusions}
\label{sec:conclusions}
%%%%%%%%%%%%%%%%%%%%%%%%%%%%%
In this work we have investigated one-parameter modular Yukawa sectors, in which a charged-fermion mass matrix is generated by a single modular-invariant contraction and is therefore controlled, up to an overall normalization, only by the modulus $\tau$. We formulated the conditions under which this can happen: the matter fields must be assigned to triplets, the modulus $\tau$ must be close to the cusp (large $\im\tau$) and the corresponding mass matrix has rank at most one in the symmetric point, that is hierarchically lifted to 3 in its vicinity. We then performed a systematic search of small finite modular groups (order $\leq 100$) that could contain \acp{OPM}, leading us to identify the modular version of the $\Delta(96)$ group as the only possible candidate at these orders. A non-exhaustive follow-up search further identified the modular version of the larger $\Delta(384)$ group as an interesting candidate.

The possible groups and representations are highly restricted by the \ac{OPM} requirement, before any phenomenological input is imposed. We found $5$ hierarchy patterns (four inequivalent) allowed by $\Delta(96)$ and $\Delta(384)$, summarized in~\Cref{tab:H12345}. \acp{OPM} do not provide a flexible parametrization of fermion masses, but rather a discrete set of predictive curves in the hierarchy plane.

Within the congruence finite-image framework considered here, $\mathrm{SL}(2,\mathbb{Z})$ is known to have only finitely many irreducible three-dimensional representations.\footnote{ \Acp{irrep} of $\mathrm{SL}(2,\mathbb{Z})$ with congruence subgroup kernels are also \acp{irrep} of homogeneous finite quotient groups $\Gamma'_N$. There are in total 144 inequivalent three-dimensional \acp{irrep} of this kind. Notably, three-dimensional \acp{irrep} of $\mathrm{SL}(2,\mathbb{Z})$ may also have non-congruence subgroup kernels (a feature absent in one-dimensional and two-dimensional \acp{irrep}), in which case there exists an infinite sequence of so-called imprimitive \acp{irrep}. However, this is not the main focus of our current one-parameter models and we leave related work for future study.} Consequently, in \acp{OPM}, the total number of possible assignments of representations for left-handed and right-handed fermion fields is finite. Moreover, the dimension of \ac{VVMF} spaces is also known to be finite~\cite{Liu:2021gwa}. In particular, when constructing \acp{OPM}, we typically have the option to choose only the lowest-weight \ac{VVMF} as the Yukawa coupling. Thus, in practice, \acp{OPM} based on $\mathrm{SL}(2,\mathbb{Z})$ modular symmetry are quite limited, yielding only a finite number of distinct prediction sets. As a result, finding a pattern that matches quark or lepton masses among these finitely many \acp{OPM} is highly non-trivial.

As an explicit proof of principle, we have constructed a double \ac{OPM} based on $\Delta(384)$, in which the charged-lepton and down-quark sectors are controlled by a common modulus. Up to corrections of very high-order in $\epsilon = e^{-\pi \im\tau/8}$, the charged-lepton sector realizes the H$_2$ hierarchy,
\[
m_e:m_\mu:m_\tau
=
4\sqrt{2}\,\epsilon^6:2\epsilon^2:1\,,
\]
while the down-quark sector realizes the H$_5$ hierarchy,
\[
m_d:m_s:m_b
=
4\epsilon^5:2\sqrt{2}\,\epsilon^3:1\,.
\]
Since the two sectors share the same expansion parameter $\epsilon$, the model predicts three exact high-scale mass relations,
\[
m_s^5=2\sqrt{2}\,m_d^3m_b^2\,,\qquad
m_\mu^3=\sqrt{2}\,m_e m_\tau^2\,,\qquad
m_s^2m_\tau=\sqrt{2}\,m_e m_b^2\,.
\]
These relations are fixed by the modular construction and do not rely on additional order-one flavour coefficients.

We then compared these high-scale relations with low-energy charged-fermion data by including the \ac{RG} evolution between the flavour and electroweak scales. Running effects alone already reveal a non-trivial alignment: both down-quark hierarchies and the electron-to-tau hierarchy can be brought into agreement with data for a common value of $\epsilon$, while the main remaining discrepancy is isolated in the muon-to-tau ratio. We showed that this discrepancy can be corrected by a selective charged-lepton SUSY threshold effect. In this model, the high-scale \ac{OPM} structure captures most of the charged-fermion hierarchy, while the remaining adjustment can be associated with a specific pattern of threshold effects.

The two one-parameter sectors can be incorporated into a complete quark model. In Appendix~\ref{app:up-completion}, we construct a weight-10 up-quark Yukawa matrix that accommodates the up-quark masses and CKM observables at the same value of the modulus. Its complex coefficients are fitted to these observables; the mass relations of the charged-lepton and down-quark sectors are preserved.

Several research directions remain open. One direction is to search for an up-quark \ac{OPM} that also accounts for CKM mixing with fewer flavour parameters than the completion of Appendix~\ref{app:up-completion}. Another is to study the soft supersymmetry-breaking sector in a more top-down way, including modular-covariant soft terms and a systematic treatment of finite threshold corrections, and to understand how the sharp mass relations derived here are modified by non-minimal Kähler effects. Finally, it would be important to investigate whether the \acp{OPM} constructed here can be realized in string theory. Finite modular groups such as $S_3$, $T'$, and $2D_3 \subset S'_4$ have been shown to arise naturally in $T^2/\mathbb{Z}_N$ heterotic orbifolds~\cite{Baur:2024qzo}. Extending this picture to larger groups such as $\Delta(96)$ and $\Delta(384)$ would provide a compelling ultraviolet origin for the present construction. The fact that lowest-weight \acp{VVMF} often appear in leading trilinear couplings in string constructions is suggestive in this respect. Clarifying these issues will determine whether \acp{OPM} can go beyond predicting charged-fermion mass relations and provide a genuine solution to the quark flavour puzzle.

\section*{Acknowledgements}
X.~L.~and X.-G.~L. would like to thank Michael Ratz and Mu-Chun Chen for useful discussions and support. X.-G.~L.~was also supported by the Universidad Nacional Autónoma de México Postdoctoral Program (POSDOC). S.~C.~Ch.~acknowledges support from the Spanish grants PID2023-147306NB-I00, CNS2024-154524 and CEX2023-001292-S (MICIU/AEI/10.13039/501100011033).

%%%%%%%%%%%%%%%%%%%%%%%
\appendix
%%%%%%%%%%%%%%%%%%%%%%%

%%%%%%%%%%%%%%%%%%%%%%%%%%%%%
\section{\acp{VVMF} and \acp{OPM}}
\label{app:A}
%%%%%%%%%%%%%%%%%%%%%%%%%%%%%

We collect here the complete list of \ac{OPM} triplet pairs found, as described in~\Cref{sec:models}. Throughout this appendix we consider $k_{\psi^c}+k_\psi=2$, as well as a trivial and weightless Higgs representation, $\rho_H\sim\mathbf{1}$ and $k_H=0$. For each \ac{OPM}, the superpotential takes the form
\begin{equation}
\mathcal{W}_f=\alpha_f\left(Y^{(2)}_{\mathbf r_Y}(\tau)\,\psi^c\psi\right)_{\mathbf 1}H\,,
\end{equation}
where the modular-form representation $\mathbf r_Y$ is specified below.

The corresponding hierarchical spectra arise in the vicinity of the cusp, with $\epsilon=|q|^{1/N}$, where $N=8$ for $\Delta(96)$ and $N=16$ for $\Delta(384)$. For $\Delta(96)$, the two triplet pairs leading to an \ac{OPM} are shown in~\Cref{tab:opm_delta96}. They both realize the same hierarchy pattern, denoted H$_1$ in~\Cref{tab:H12345}. For $\Delta(384)$, the eight triplet pairs found in the search are given in~\Cref{tab:opm_delta384}. They come in four pairs, each leading to one of the four leading-order hierarchy patterns $\text{H}_2,\dots,\text{H}_5$, see~\Cref{tab:H12345}. The two $\Delta(384)$ models realizing H$_2$ reproduce the same hierarchy as the $\Delta(96)$ models realizing H$_1$. The pairwise degeneracy of the \acp{OPM} observed in~\Cref{tab:opm_delta96} and ~\Cref{tab:opm_delta384} is a direct manifestation of the outer automorphisms of the finite modular groups. Specifically, there exists a non-trivial outer automorphism that permutes the triplet irreducible representations, such as $(\mathbf{3}_0, \mathbf{3}_9) \leftrightarrow (\mathbf{3}_1, \mathbf{3}_2)$, while leaving the sextet multiplets $\mathbf{6}_0$ invariant. Since the paired models share the exact same sextet \ac{VVMF} $Y^{(2)}_{\mathbf 6_0}$ for their Yukawa couplings, this algebraic symmetry ensures that their respective \ac{CG} contractions are structurally isomorphic. Consequently, the resulting mass matrices and the predicted mass hierarchies remain physically equivalent for each pair.

\begin{table}[t]
\centering
\renewcommand{\arraystretch}{1.25}
\begin{tabular}{c c c c c}
\toprule
\quad{ }Model\quad{ } & \quad{ }$(\rho_{\psi^c},\rho_\psi)$\quad{ } & \quad{ }$\mathbf r_Y$\quad{ } & \quad{ }$m_1:m_2:m_3$\quad{ } & \quad{ }Hierarchy\quad{ }\\
\midrule
$\Delta(96)$-1 & $(\mathbf 3_0,\mathbf 3_5)$ & $\mathbf 6$ & $4\sqrt{2}\,\epsilon^3:2\epsilon:1$ & H$_1$\\
$\Delta(96)$-2 & $(\mathbf 3_1,\mathbf 3_2)$ & $\mathbf 6$ & $4\sqrt{2}\,\epsilon^3:2\epsilon:1$ & H$_1$\\
\bottomrule
\end{tabular}
\caption{Triplet pairs leading to one-parameter models based on $\Delta(96)$. Here $\epsilon=|q|^{1/8}$.}
\label{tab:opm_delta96}
\end{table}

\begin{table}[t]
\centering
\renewcommand{\arraystretch}{1.25}
\begin{tabular}{c c c c c}
\toprule
\quad{ }Model\quad{ } & \quad{ }$(\rho_{\psi^c},\rho_\psi)$\quad{ } & \quad{ }$\mathbf r_Y$\quad{ } & \quad{ }$m_1:m_2:m_3$\quad{ } & \quad{ }Hierarchy\quad{ }\\
\midrule
$\Delta(384)$-1 & $(\mathbf 3_0,\mathbf 3_9)$ & $\mathbf 6_0$ & $4\sqrt{2}\,\epsilon^6:2\epsilon^2:1$ & H$_2$\\
$\Delta(384)$-2 & $(\mathbf 3_1,\mathbf 3_2)$ & $\mathbf 6_0$ & $4\sqrt{2}\,\epsilon^6:2\epsilon^2:1$ & H$_2$\\
$\Delta(384)$-3 & $(\mathbf 3_0,\mathbf 3_{13})$ & $\mathbf 6_5$ & $2\sqrt{2}\,\epsilon^4:\sqrt{2}\epsilon:1$ & H$_3$\\
$\Delta(384)$-4 & $(\mathbf 3_1,\mathbf 3_6)$ & $\mathbf 6_5$ & $2\sqrt{2}\,\epsilon^4:\sqrt{2}\epsilon:1$ & H$_3$\\
$\Delta(384)$-5 & $(\mathbf 3_2,\mathbf 3_{10})$ & $\mathbf 6_3$ & $8\epsilon^7:\sqrt{2}\epsilon:1$ & H$_4$\\
$\Delta(384)$-6 & $(\mathbf 3_5,\mathbf 3_9)$ & $\mathbf 6_3$ & $8\epsilon^7:\sqrt{2}\epsilon:1$ & H$_4$\\
$\Delta(384)$-7 & $(\mathbf 3_3,\mathbf 3_{12})$ & $\mathbf 6_4$ & $4\epsilon^5:2\sqrt{2}\epsilon^3:1$ & H$_5$\\
$\Delta(384)$-8 & $(\mathbf 3_7,\mathbf 3_8)$ & $\mathbf 6_4$ & $4\epsilon^5:2\sqrt{2}\epsilon^3:1$ & H$_5$\\
\bottomrule
\end{tabular}
\caption{Triplet pairs leading to one-parameter models based on $\Delta(384)$. Here $\epsilon=|q|^{1/16}$. For the H$_3$ entries, a common overall factor of $\epsilon$ has been factored out.}
\label{tab:opm_delta384}
\end{table}

The leading behaviour of the relevant $\Delta(384)$ sextet \acp{VVMF} close to the cusp is
\begin{align}
Y^{(2)}_{\mathbf 6_0}(\tau)\sim\begin{pmatrix} -8\,\epsilon^6\\2\,\epsilon^2\\ -4\,\epsilon^8\\ 1 \\-12\,\epsilon^{10}\\ -16\,\epsilon^{14}\end{pmatrix}\,,\quad
Y^{(2)}_{\mathbf 6_5}(\tau)\sim\begin{pmatrix}-\sqrt{2}\,\epsilon^2\\3\,\epsilon^9\\\epsilon\\-4\sqrt{2}\,\epsilon^{10}\\ 4\,\epsilon^5\\ -4\,\epsilon^{13}\end{pmatrix}\,,\quad
Y^{(2)}_{\mathbf 6_3}(\tau)\sim\begin{pmatrix} -8\sqrt{2}\,\epsilon^7\\8\sqrt{2}\,\epsilon^{15}\\-6\,\epsilon^8\\1\\-7\sqrt{2}\,\epsilon^9\\\sqrt{2}\,\epsilon\end{pmatrix}\,,\quad
Y^{(2)}_{\mathbf 6_4}(\tau)\sim\begin{pmatrix}10\sqrt{2}\,\epsilon^{11}\\2\sqrt{2}\,\epsilon^3\\-2\,\epsilon^8\\-1\\4\sqrt{2}\,\epsilon^5\\-12\sqrt{2}\,\epsilon^{13}\end{pmatrix},
\end{align}
up to ($\re\tau$)-dependent phases and higher-order corrections in $\epsilon$. These leading powers, combined with the \ac{CG} contractions for the triplet pairs in~\Cref{tab:opm_delta384}, up to transposition and permutations and sign flips of rows and columns, result in the Yukawa matrices:
\begin{subequations}
\begin{align}
\mathcal{Y}_2(\tau) &\,\propto\,
\begin{pmatrix}
0 & \sqrt{2}\,Y^{(2)}_{\mathbf{6}_0,5} & -\sqrt{2}\,Y^{(2)}_{\mathbf{6}_0,2} \\
\sqrt{2}\,Y^{(2)}_{\mathbf{6}_0,4} & Y^{(2)}_{\mathbf{6}_0,6} & Y^{(2)}_{\mathbf{6}_0,1} \\
-\sqrt{2}\,Y^{(2)}_{\mathbf{6}_0,3} & -Y^{(2)}_{\mathbf{6}_0,1} & -Y^{(2)}_{\mathbf{6}_0,6}
\end{pmatrix}
\,\,\, \sim \,\,\,
\begin{pmatrix}
0 & -12\sqrt{2}\,\epsilon^{10} & -2\sqrt{2}\,\epsilon^2 \\
\sqrt{2} & -16\,\epsilon^{14} & -8\,\epsilon^6 \\
4\sqrt{2}\,\epsilon^8 & 8\,\epsilon^6 & 16\,\epsilon^{14}
\end{pmatrix}
\,,
\\[2mm]
\mathcal{Y}_3(\tau) &\,\propto\,
\begin{pmatrix}
\sqrt{2}\,Y^{(2)}_{\mathbf{6}_5,3} & -\sqrt{2}\,Y^{(2)}_{\mathbf{6}_5,2} & 0 \\
Y^{(2)}_{\mathbf{6}_5,5} & -Y^{(2)}_{\mathbf{6}_5,6} & -\sqrt{2}\,Y^{(2)}_{\mathbf{6}_5,1} \\
-Y^{(2)}_{\mathbf{6}_5,6} & Y^{(2)}_{\mathbf{6}_5,5} & -\sqrt{2}\,Y^{(2)}_{\mathbf{6}_5,4}
\end{pmatrix}
\,\,\, \sim \,\,\,
\begin{pmatrix}
\sqrt{2}\,\epsilon & -3\sqrt{2}\,\epsilon^9 & 0 \\
4\,\epsilon^5 & 4\,\epsilon^{13} & 2\,\epsilon^2 \\
4\,\epsilon^{13} & 4\,\epsilon^5 & 8\,\epsilon^{10}
\end{pmatrix}
\,,
\\[2mm]
\mathcal{Y}_4(\tau) &\,\propto\,
\begin{pmatrix}
\sqrt{2}\,Y^{(2)}_{\mathbf{6}_3,5} & \sqrt{2}\,Y^{(2)}_{\mathbf{6}_3,6} & 0 \\
Y^{(2)}_{\mathbf{6}_3,1} & Y^{(2)}_{\mathbf{6}_3,2} & -\sqrt{2}\,Y^{(2)}_{\mathbf{6}_3,3} \\
Y^{(2)}_{\mathbf{6}_3,2} & Y^{(2)}_{\mathbf{6}_3,1} & \sqrt{2}\,Y^{(2)}_{\mathbf{6}_3,4}
\end{pmatrix}
\,\,\, \sim \,\,\,
\begin{pmatrix}
-14\,\epsilon^9 & 2\,\epsilon & 0 \\
-8\sqrt{2}\,\epsilon^7 & 8\sqrt{2}\,\epsilon^{15} &6\sqrt{2}\,\epsilon^8 \\
8\sqrt{2}\,\epsilon^{15} & -8\sqrt{2}\,\epsilon^7 & \sqrt{2}
\end{pmatrix}
\,,
\\[2mm]
\mathcal{Y}_5(\tau) &\,\propto\,
\begin{pmatrix}
-\sqrt{2}\,Y^{(2)}_{\mathbf{6}_4,3} & 0 & \sqrt{2}\,Y^{(2)}_{\mathbf{6}_4,4} \\
Y^{(2)}_{\mathbf{6}_4,5} & -\sqrt{2}\,Y^{(2)}_{\mathbf{6}_4,1} & -Y^{(2)}_{\mathbf{6}_4,6} \\
-Y^{(2)}_{\mathbf{6}_4,6} & -\sqrt{2}\,Y^{(2)}_{\mathbf{6}_4,2} & Y^{(2)}_{\mathbf{6}_4,5}
\end{pmatrix}
\,\,\, \sim \,\,\,
\begin{pmatrix}
2\sqrt{2}\,\epsilon^8 & 0 & -\sqrt{2} \\
4\sqrt{2}\,\epsilon^5 & -20\,\epsilon^{11} & 12\sqrt{2}\,\epsilon^{13} \\
12\sqrt{2}\,\epsilon^{13} & -4\,\epsilon^3 & 4\sqrt{2}\,\epsilon^5
\end{pmatrix}
\,,
\end{align}
\end{subequations}
where we have indicated the $\epsilon$ power structure for each matrix, which produce the hierarchy patterns $\text{H}_2,\dots,\text{H}_5$.

In the phenomenological construction discussed in the main text, only the H$_2$ and H$_5$ models are used explicitly. We therefore give below the analytical expressions in terms of theta constants and the corresponding $q$-expansions for the two sextet forms entering that construction, $Y^{(2)}_{\mathbf 6_0}$ and $Y^{(2)}_{\mathbf 6_4}$.

The relevant $\Delta(384)$ \acp{VVMF} can be constructed from weight-$1/2$ modular forms on $\Gamma(8)$ and $\Gamma(16)$. In the convention used here, these building blocks are the theta constants\footnote{It should be noted that these theta constants are not linearly independent. As can be easily shown, $\theta_2=\theta_4$, $\vartheta_2=\vartheta_8$, $\vartheta_3=\vartheta_7$, and $\vartheta_4=\vartheta_6$.}
\begin{subequations}
\begin{align}
\theta_i(\tau)&\equiv\sum_{n=-\infty}^{\infty}e^{4\pi i\tau(n+\frac{i-1}{4})^2}\,,\qquad i=1,\dots,4\,,\\
\vartheta_i(\tau)&\equiv\sum_{n=-\infty}^{\infty}e^{8\pi i\tau(n+\frac{i-1}{8})^2}\,,\qquad i=1,\dots,8\,.
\end{align}
\end{subequations}
In terms of these theta constants, the two sextet modular forms used in the explicit $\text{H}_2+\text{H}_5$ construction are
\begin{subequations}
\begin{align}
Y^{(2)}_{\mathbf{6}_0}&=\left(
\begin{array}{c}
 -4 \left(\vartheta _1+\vartheta _5\right) \left(\vartheta _3+\vartheta _7\right) \left(\vartheta _4 \vartheta _6+\vartheta _2 \vartheta _8\right) \\
 \left(\vartheta _2^2+\vartheta _4^2+\vartheta _6^2+\vartheta _8^2\right) \left(\vartheta _1+\vartheta _5\right){}^2+2 \left(\vartheta _3+\vartheta _7\right){}^2 \left(\vartheta _2 \vartheta _6+\vartheta _4 \vartheta _8\right) \\
 4 \left(\vartheta _3 \vartheta _5-\vartheta _1 \vartheta _7\right) \left(\vartheta _1 \vartheta _3-\vartheta _5 \vartheta _7\right) \\
 \left(\vartheta _1^2-\vartheta _5^2\right){}^2-\left(\vartheta _3^2-\vartheta _7^2\right){}^2 \\
 -2 \left(\vartheta _2 \vartheta _6+\vartheta _4 \vartheta _8\right) \left(\vartheta _1+\vartheta _5\right){}^2-\left(\vartheta _3+\vartheta _7\right){}^2 \left(\vartheta _2^2+\vartheta _4^2+\vartheta _6^2+\vartheta _8^2\right) \\
 -4 \left(\vartheta _1+\vartheta _5\right) \left(\vartheta _3+\vartheta _7\right) \left(\vartheta _2 \vartheta _4+\vartheta _6 \vartheta _8\right)
\end{array}
\right)\,,\\
Y^{(2)}_{\mathbf{6}_4}&=\left(
\begin{array}{c}
 \sqrt{2} \left(\theta _1^2 \left(\theta _2 \vartheta _4+\theta _4 \vartheta _6\right)+\theta _3 \left(\theta _2+\theta _4\right) \theta _1\left(\vartheta _2+\vartheta _8\right)+\theta _3^2 \left(\theta _4 \vartheta _4+\theta _2 \vartheta _6\right)\right) \\
 \sqrt{2} \left(\theta _2 \left(\theta _1^2 \vartheta _8+\theta _3 \theta _1 \left(\vartheta _4+\vartheta _6\right)+\theta _3^2 \vartheta _2\right)+\theta _4 \left(\theta _1^2 \vartheta _2+\theta _3 \theta _1 \left(\vartheta _4+\vartheta _6\right)+\theta _3^2 \vartheta _8\right)\right) \\
 \left(\theta _3^3-\theta _1^2 \theta _3\right) \left(\vartheta _1-\vartheta _5\right)-\theta _1 \left(\theta _2^2-\theta _4^2\right) \left(\vartheta _3-\vartheta _7\right) \\
 \left(\theta _1 \theta _3^2-\theta _1^3\right) \left(\vartheta _1-\vartheta _5\right)-\theta _3 \left(\theta _2^2-\theta _4^2\right) \left(\vartheta _3-\vartheta _7\right) \\
 \sqrt{2} \left(\theta _2+\theta _4\right) \left(\theta _3 \left(\theta _2 \vartheta _4+\theta _4 \vartheta _6\right)+\theta _1 \left(\theta _2 \vartheta _2+\theta _4 \vartheta _8\right)\right) \\
 -\sqrt{2} \left(\theta _2+\theta _4\right) \left(\theta _1 \left(\theta _4 \vartheta _4+\theta _2 \vartheta _6\right)+\theta _3 \left(\theta _4 \vartheta _2+\theta _2 \vartheta _8\right)\right)
\end{array}
\right)\,.
\end{align}
\end{subequations}
Expanding around the cusp, with $q\equiv e^{2\pi i\tau}$, one obtains
\begin{equation}
Y^{(2)}_{\mathbf{6}_0}(\tau)=\begin{pmatrix}
 -8 q^{3/8} (1 + 3 q + 5 q^2 + 10 q^3 + 12 q^4 + 11 q^5 + 18 q^6 + 15 q^7 + 17 q^8 + 31 q^9 + 21 q^{10} +\dots) \\
 2 q^{1/8} (1 + 13 q + 18 q^2 + 31 q^3 + 48 q^4 + 42 q^5 + 57 q^6 + 80 q^7 + 84 q^8 + 74 q^9+ 121 q^{10}+\dots) \\
 4 q^{1/2}(-1 + 4 q - 6 q^2 + 8 q^3 - 13 q^4 + 12 q^5 - 14 q^6 + 24 q^7 - 18 q^8 + 20 q^9 - 32 q^{10} +\dots) \\
 1 - 8 q^2 + 24 q^4 - 32 q^6 + 24 q^8 - 48 q^{10}+\dots \\
 -4 q^{5/8} (3 + 7 q + 16 q^2 + 15 q^3 + 19 q^4 + 39 q^5 + 27 q^6 + 31 q^7 + 48 q^8 + 48 q^9 + 54 q^{10} +\dots)\\
 -16 q^{7/8} (1 + 3 q + 3 q^2 + 4 q^3 + 7 q^4 + 6 q^5 + 9 q^6 + 13 q^7 + 9 q^8 + 10 q^9 + 15 q^{10}+\dots)
\end{pmatrix}\,,
\label{eq:Y60_weight2}
\end{equation}
and
\begin{equation}
Y^{(2)}_{\mathbf{6}_4}(\tau)=\begin{pmatrix}
 2\sqrt{2} q^{11/16} (5 + 10 q + 21 q^2 + 29 q^3 + 21 q^4 + 48 q^5 + 53 q^6 + 42 q^7 + 69 q^8 + 64 q^9 + 63 q^{10} +\dots) \\
 2\sqrt{2} q^{3/16} (1 + 9 q + 16 q^2 + 18 q^3 + 33 q^4 + 41 q^5 + 35 q^6 + 48 q^7 + 65 q^8 + 57 q^9 + 81 q^{10}+\dots) \\
 -2q^{1/2}(1 - 6 q + 12 q^2 - 8 q^3 + 7 q^4 - 30 q^5 + 36 q^6 - 8 q^7 + 18 q^8 - 54 q^9 + 48 q^{10} +\dots) \\
 -1 + 6 q - 14 q^2 + 20 q^3 - 30 q^4 + 40 q^5 - 36 q^6 + 48 q^7 - 62 q^8 + 42 q^9 - 72 q^{10} +\dots \\
 4\sqrt{2} q^{5/16} (1 + 4 q + 9 q^2 + 13 q^3 + 12 q^4 + 18 q^5 + 25 q^6 + 21 q^7 + 36 q^8 + 37 q^9 + 20 q^{10} +\dots)\\
 -4\sqrt{2} q^{13/16} (3 + 7 q + 7 q^2 + 15 q^3 + 20 q^4 + 16 q^5 + 27 q^6 + 26 q^7 + 24 q^8 + 39 q^9 + 43q^{10}+\dots)
\end{pmatrix}\,.
\label{eq:Y64_weight2}
\end{equation}

%%%%%%%%%%%%%%%%%%%%%%%%%%%%%%%%%%%%%%%%%%%%%%%%%%%%%%%%%%
\section{A non-minimal completion of the up-quark sector}
\label{app:up-completion}
%%%%%%%%%%%%%%%%%%%%%%%%%%%%%%%%%%%%%%%%%%%%%%%%%%%%%%%%%%

We present an explicit completion of the up-quark sector for the $\Delta(384)$ construction of~\Cref{sec:models}. The assignments $(\rho_{d^c},\rho_Q)\sim(\mathbf{3}_7,\mathbf{3}_8)$ and $k_{d^c}+k_Q=2$ are retained, so the down-quark Yukawa matrix remains the one given in~\cref{eq:Md}. We take $H_u$ to be a trivial representation with vanishing modular weight. We take
\begin{equation}
 \rho_{u^c} \sim \mathbf{3}_0\,,\qquad
 \rho_Q \sim \mathbf{3}_8\,,\qquad
 k_{u^c}+k_Q=10\,.
 \label{eq:up_rep_assignment}
\end{equation}
Only the sums of matter-field weights enter the Yukawa operators. One possible assignment consistent with both quark sectors is $k_Q=0$, $k_{d^c}=2$, and $k_{u^c}=10$. The relevant decomposition is
\begin{equation}
\mathbf{3}_0\otimes\mathbf{3}_8
\,=\, \mathbf{3}_2 \oplus \mathbf{6}_0 = \overline{\mathbf{3}}_3 \oplus \overline{\mathbf{6}}_0 \,.
\end{equation}
The modular forms required for invariant contractions must then transform as a $\mathbf{3}_3$ triplet and a $\mathbf{6}_0$ sextet.

In the basis used for the contractions below, the generators $\rho(T)$ are diagonal. Each of the considered \acp{irrep} can be uniquely identified by:
\begin{equation} \label{eq:up_T_generators}
\frac{1}{2\pi i}\log(\rho_\mathbf{r})(T)\,\,\,\,\text{mod}\,\mathbb{Z}\, =\,
\begin{cases}
\diag\left(\frac{1}{4},\,\frac{3}{4},\,0\right)\,,&\text{for }\mathbf{r} = \mathbf{3}_0\,,\\
\diag\left(\frac{3}{4},\,\frac{7}{8},\,\frac{3}{8}\right)\,,&\text{for }\mathbf{r} = \mathbf{3}_2\,,\\
\diag\left(\frac{1}{4},\,\frac{5}{8},\,\frac{1}{8}\right)\,,&\text{for }\mathbf{r} = \mathbf{3}_3\,,\\
\diag\left(\frac{1}{8},\,\frac{3}{4},\,\frac{5}{8}\right)\,,&\text{for }\mathbf{r} = \mathbf{3}_8\,,\\
\diag\left(\frac{3}{8},\,\frac{1}{8},\,\frac{1}{2},\,0,\,\frac{5}{8},\,\frac{7}{8}\right)\,,&\text{for }\mathbf{r} = \mathbf{6}_0\,.
\end{cases}
\end{equation}
The sextet eigenvalues are listed in the order of the components in~\cref{eq:Y60_weight2}.

The free-module generator weights are $(2,4,6)$ for $\mathbf{3}_3$ and $(2,4,4,6,6,8)$ for $\mathbf{6}_0$. At weight 10 these give, respectively, three and five independent vector-valued modular forms. A basis for the triplet forms can be chosen as
\begin{equation} \label{eq:up_weight10_triplets}
 Y^{(10)}_{\mathbf{3}_3,a}\,=\,E_4^2\,Y^{(2)}_{\mathbf{3}_3}\,, \quad
 Y^{(10)}_{\mathbf{3}_3,b}\,=\,E_6\,Y^{(4)}_{\mathbf{3}_3}\,, \quad
 Y^{(10)}_{\mathbf{3}_3,c}\,=\,E_4\,Y^{(6)}_{\mathbf{3}_3,b}\,,
\end{equation}
while a basis for the sextet forms is
\begin{equation} \label{eq:up_weight10_sextets}
\begin{aligned}
&Y^{(10)}_{\mathbf{6}_0,a}\,=\,E_4^2\,Y^{(2)}_{\mathbf{6}_0}\,, \quad
 Y^{(10)}_{\mathbf{6}_0,b}\,=\,E_6\,Y^{(4)}_{\mathbf{6}_0,a}\,, \quad
 Y^{(10)}_{\mathbf{6}_0,c}\,=\,E_6\,Y^{(4)}_{\mathbf{6}_0,b}\,,\\
&Y^{(10)}_{\mathbf{6}_0,d}\,=\,E_4\,Y^{(6)}_{\mathbf{6}_0,b}\,, \quad
 Y^{(10)}_{\mathbf{6}_0,e}\,=\,E_4\,Y^{(6)}_{\mathbf{6}_0,c}\,.
\end{aligned}
\end{equation}
Here, the letter subscripts distinguish independent forms of the same weight and representation. The lower-weight forms appearing in~\cref{eq:up_weight10_triplets,eq:up_weight10_sextets} can be constructed from the weight-2 generators by modular differentiation and tensor products.

We give the invariant contractions explicitly. For a triplet $y=(y_1,y_2,y_3)^T \sim \mathbf{3}_3$ and a sextet $z=(z_1,\ldots,z_6)^T \sim \mathbf{6}_0$, their contributions to the Yukawa matrix, in the right-left convention of the main text, are
\begin{equation}
 \mathcal C_{\mathbf{3}_3}(y)=
 \begin{pmatrix}
 -y_2/\sqrt{2}&0&y_3/\sqrt{2}\\
 y_3/\sqrt{2}&0&-y_2/\sqrt{2}\\
 0&y_1&0
 \end{pmatrix},
 \qquad
 \mathcal C_{\mathbf{6}_0}(z)=
 \begin{pmatrix}
 z_5/\sqrt{2}&-z_4&z_2/\sqrt{2}\\
 -z_2/\sqrt{2}&z_3&-z_5/\sqrt{2}\\
 z_6&0&-z_1
 \end{pmatrix}\,.
 \label{eq:up_contractions}
\end{equation}
The resulting up-quark Yukawa matrix is
\begin{equation} \label{eq:up_Yukawa_completion}
\begin{aligned}
 Y_u(\tau)={}&
 c_1\,\mathcal C_{\mathbf{3}_3}\!\left(Y^{(10)}_{\mathbf{3}_3,a}\right)
 +c_2\,\mathcal C_{\mathbf{3}_3}\!\left(Y^{(10)}_{\mathbf{3}_3,b}\right)
 +c_3\,\mathcal C_{\mathbf{3}_3}\!\left(Y^{(10)}_{\mathbf{3}_3,c}\right)
 \\
 &+c_4\,\mathcal C_{\mathbf{6}_0}\!\left(Y^{(10)}_{\mathbf{6}_0,a}\right)
 +c_5\,\mathcal C_{\mathbf{6}_0}\!\left(Y^{(10)}_{\mathbf{6}_0,b}\right)
 +c_6\,\mathcal C_{\mathbf{6}_0}\!\left(Y^{(10)}_{\mathbf{6}_0,c}\right)
 \\
 &+c_7\,\mathcal C_{\mathbf{6}_0}\!\left(Y^{(10)}_{\mathbf{6}_0,d}\right)
 +c_8\,\mathcal C_{\mathbf{6}_0}\!\left(Y^{(10)}_{\mathbf{6}_0,e}\right)\,,
\end{aligned}
\end{equation}
where the $c_i$ are complex coefficients and $M_u=v_uY_u$, with $v_u=\langle H_u^0\rangle$. Henceforth, we denote by $\mathcal M_i(\tau)$ each of the eight matrices multiplying the $c_i$ in~\cref{eq:up_Yukawa_completion}.

We now show that the up-quark masses and CKM matrix can be accommodated at the value of the modulus selected by the down-quark and charged-lepton sectors. At $\epsilon=0.1892$ and $\re\tau=0$, corresponding to $\tau\simeq4.24\,i$, direct evaluation gives
\begin{equation}
 \dim_{\mathbb C}\operatorname{span}
 \bigl\{\mathcal M_1(\tau),\ldots,\mathcal M_8(\tau)\bigr\}=8\,.
 \label{eq:up_matrix_rank}
\end{equation}
Thus these matrices span a codimension-one subspace of the set of $3\times 3$ complex matrices $M_3(\mathbb C)$.

Let $V_{dL}$ diagonalize $Y_d^\dagger Y_d$ and define $\widehat{\mathcal M}_i=\mathcal M_iV_{dL}$. Right multiplication by $V_{dL}$ preserves their linear independence. There is consequently a non-zero matrix $N$, unique up to an overall complex factor, such that
\begin{equation}
 \tr
 \bigl(N^\dagger\widehat{\mathcal M}_i\bigr)=0\,,
 \qquad i=1,\ldots,8\,.
 \label{eq:up_orthogonal_complement}
\end{equation}
In such a down-diagonal basis, a Yukawa matrix leading to the desired up-quark singular values and CKM matrix data has the form
\begin{equation}
 \widehat Y_u^{\mathrm{phys}}
 =U_R D_u V_{\mathrm{CKM}}\,,
 \qquad
 D_u=\diag(y_u,y_c,y_t)\,,
 \qquad U_R\in U(3)\,.
 \label{eq:up_target_matrix}
\end{equation}
The unitary matrix $U_R$ acts on the up-type singlets and is unconstrained by quark masses and mixing. The target matrix $\widehat Y_u^{\mathrm{phys}}$ belongs to the span in~\cref{eq:up_matrix_rank} if
\begin{equation} \label{eq:up_span_condition}
 \tr \bigl(N^\dagger U_RD_uV_{\mathrm{CKM}}\bigr)=0\,.
\end{equation}
Such a choice of $U_R$ can be constructed directly. Define $A\equiv D_uV_{\mathrm{CKM}}N^\dagger$ and write its singular-value decomposition as $A=L\Sigma R^\dagger$, where $\Sigma$ is diagonal and $L$ and $R$ are unitary. For any cyclic permutation matrix $P$ with vanishing diagonal, take $U_R=RPL^\dagger$ (still unitary). Then
\begin{equation}
 \tr
 \bigl(N^\dagger U_RD_uV_{\mathrm{CKM}}\bigr)
 =\tr(AU_R)
 =\tr(\Sigma P)
 =0\,.
 \label{eq:up_constructive_fit}
\end{equation}
\Cref{eq:up_matrix_rank} then guarantees the existence of complex coefficients $c_i$ satisfying
\begin{equation}
 \widehat Y_u^{\mathrm{phys}} = U_RD_uV_{\mathrm{CKM}}
 =\sum_{i=1}^{8}c_i\widehat{\mathcal M}_i\,.
 \label{eq:up_fit_coefficients}
\end{equation}
Hence the weight-10 completion can reproduce the up-quark masses and CKM observables at the same modulus value used in the one-parameter down-quark and charged-lepton sectors.

The eight complex coefficients $c_i$ contain 15 physical real parameters, since one overall phase can be removed by a field rephasing. They accommodate seven observables: the three up-quark Yukawa couplings, the three CKM mixing angles and the CP violation phase. The completion therefore makes no prediction for the up-quark masses or for CKM mixing. Note that the construction of $U_R$ above applies to any $D_u$ and any unitary $V_{\mathrm{CKM}}$; in particular, it applies to the up-quark data at the flavour scale. At $\re\tau=0$ the matrices $\mathcal M_i(\tau)$ are real, so CP violation in the quark sector arises only from the phases of the $c_i$ and has no modular origin in this completion. For this assignment of representations, weight 10 is the lowest weight at which the up-quark data can be accommodated: at weights 2, 4, 6 and 8 there are 2, 3, 5 and 6 independent contractions, respectively, and we find numerically that no choice of coefficients reproduces the up-quark masses together with the CKM matrix. The mass relations of the down-quark and charged-lepton sectors are unaffected. The H$_6$ up-quark hierarchy considered in Section~\ref{sec:quark-flavour-hints} remains a separate, more predictive (but speculative) possibility.

%%%%%%%%%%%%%%%%%%%%%%%%%%%%%
\bibliographystyle{JHEPwithnote}
\bibliography{references}
%%%%%%%%%%%%%%%%%%%%%%%

\end{document}